\documentclass[twocolumn]{aastex631}

\usepackage[caption=false]{subfig}

\usepackage{amsmath}

\begin{document}

\title{The Atacama Cosmology Telescope: Millimeter Observations of a Population of Asteroids\\or: ACTeroids}


\author[0000-0003-1842-8104]{John Orlowski-Scherer}
\affiliation{Department of Physics, McGill University, 3600 Rue University, Montr\'{e}al, QC, H3A 2T8, Canada}

\author[0000-0003-3299-3804]{Ricco C. Venterea}
\affiliation{Department of Astronomy, Cornell University, Ithaca, NY 14853, USA}

\author[0000-0001-5846-0411]{Nicholas Battaglia}
\affiliation{Department of Astronomy, Cornell University, Ithaca, NY 14853, USA}

\author[0000-0002-4478-7111]{Sigurd Naess}
\affiliation{Institute of Theoretical Astrophysics, University of Oslo, Norway}

\author[0000-0002-2971-1776]{Tanay Bhandarkar}
\affiliation{Department of Physics and Astronomy, University of Pennsylvania, 209 South 33rd St, Philadelphia, PA, 19104, USA}

\author[0000-0002-2840-9794]{Emily Biermann}
\affiliation{Department of Physics and Astronomy, University of Pittsburgh, Pittsburgh, PA, 15213, USA}

\author[0000-0003-0837-0068]{Erminia Calabrese}
\affiliation{School of Physics and Astronomy, Cardiff University, The Parade, Cardiff, Wales CF24 3AA, UK}

\author[0000-0002-3169-9761]{Mark Devlin}
\affiliation{Department of Physics and Astronomy, University of Pennsylvania, 209 South 33rd St, Philadelphia, PA, 19104, USA}

\author[0000-0002-7450-2586]{Jo Dunkley}
\affiliation{Joseph Henry Laboratories of Physics, Jadwin Hall, Princeton University, Princeton, NJ 08544, USA}
\affiliation{Department of Astrophysical Sciences, Peyton Hall, Princeton University, Princeton, NJ USA 08544}

\author[0000-0002-4765-3426]{Carlos~Herv\'ias-Caimapo}
\affiliation{Instituto de Astrof\'isica and Centro de Astro-Ingenier\'ia, Facultad de F\'isica, Pontificia Universidad Cat\'olica de Chile, Av. Vicu\~na Mackenna 4860, 7820436 Macul, Santiago, Chile}

\author[0000-0001-9731-3617]{Patricio A. Gallardo}
\affiliation{Kavli Institute for Cosmological Physics, University of Chicago, Chicago, IL 60637, USA}
 
\author[0000-0002-8490-8117]{Matt Hilton}
\affiliation{Wits Centre for Astrophysics, School of Physics, University of the
Witwatersrand, Private Bag 3, 2050, Johannesburg, South Africa}
\affiliation{Astrophysics Research Centre, School of Mathematics, Statistics, and
Computer Science, University of KwaZulu-Natal, Westville Campus, Durban
4041, South Africa}

\author[0000-0003-1690-6678]{Adam D. Hincks}
\affiliation{David A. Dunlap Department of Astronomy and Astrophysics, University of Toronto, 50 St. George St., Toronto ON M5S 3H4, Canada}
\affiliation{Specola Vaticana (Vatican Observatory), V-00120, Vatican City State}

\author[0000-0002-8452-0825]{Kenda Knowles}
\affiliation{Centre for Radio Astronomy Techniques and Technologies, Department of Physics and Electronics, Rhodes University, P.O. Box 94, Makhanda 6140, South Africa}
\affiliation{South African Radio Astronomy Observatory, 2 Fir Street, Observatory 7925, South Africa}

\author[0000-0001-8093-2534]{Yaqiong Li}
\affiliation{Department of Physics, Cornell University, Ithaca, NY 14853, USA}
\affiliation{Kavli Institute at Cornell for Nanoscale Science, Cornell University, Ithaca, NY 14853, USA}

\author{Jeffrey J McMahon}
\affiliation{Department of Astronomy and Astrophysics, University of Chicago, Chicago, IL 60637, USA }
\affiliation{Kavli Institute for Cosmological Physics, University of Chicago, Chicago, IL 60637, USA}
\affiliation{Department of Physics, University of Chicago, Chicago, IL 60637, USA}
\affiliation{Enrico Fermi Institute, University of Chicago, Chicago, IL 60637, USA}

\author[0000-0001-7125-3580]{Michael D. Niemack}
\affiliation{Department of Physics, Cornell University, Ithaca, NY 14853, USA}
\affiliation{Department of Astronomy, Cornell University, Ithaca, NY 14853, USA}

\author[0000-0002-9828-3525]{ Lyman A. Page}
\affiliation{Joseph Henry Laboratories of Physics, Jadwin Hall, Princeton University, Princeton, NJ 08544, USA}

\author[0000-0001-6541-9265]{Bruce Partridge}
\affiliation{Department of Physics and Astronomy, Haverford College, 370 Lancaster Ave, Haverford, PA 19041, USA}

\author[0000-0003-4006-1134]{Maria~Salatino} 
\affiliation{Department of Physics, Stanford University, Stanford 94305 CA USA}\affiliation{Kavli Institute for Particle Astrophysics and Cosmology, Stanford CA, 94305 USA}

\author[0000-0001-6903-5074]{ Jonathan Sievers}
\affiliation{Department of Physics, McGill University, 3600 Rue University, Montr\'{e}al, QC, H3A 2T8, Canada}

\author[0000-0002-8149-1352]{Crist\'obal Sif\'on}
\affiliation{Instituto de F\'isica, Pontificia Universidad Cat\'olica de Valpara\'iso, Casilla 4059, Valpara\'iso, Chile}

\author[0000-0002-7020-7301]{Suzanne Staggs}
\affiliation{Joseph Henry Laboratories of Physics, Jadwin Hall, Princeton University, Princeton, NJ 08544, USA}

\author[0000-0002-3495-158X]{Alexander van Engelen}
\affiliation{School of Earth and Space Exploration, Arizona State University, Tempe, AZ 85287, USA}

\author[0000-0001-5327-1400]{Cristian~Vargas} 
\affiliation{Instituto de Astrof\'isica and Centro de Astro-Ingenier\'ia, Facultad de F\'isica, Pontificia Universidad Cat\'olica de Chile, Av. Vicu\~na Mackenna 4860, 7820436 Macul, Santiago, Chile}

\author[0000-0002-2105-7589]{Eve M. Vavagiakis}
\affiliation{Department of Physics, Cornell University, Ithaca, NY 14853, USA}

\author[0000-0002-7567-4451]{Edward J. Wollack}
\affiliation{NASA Goddard Space Flight Center, 8800 Greenbelt Road, Greenbelt, MD 20771 USA}

\begin{abstract}

We present fluxes and light curves for a population of asteroids at millimeter (mm) wavelengths, detected by the Atacama Cosmology Telescope (ACT) over $18,000$ deg$^2$ of the sky using data from 2017 to 2021. We utilize high cadence maps, which can be used in searching for moving objects such as asteroids and trans-Neptunian Objects (TNOs), as well as for studying transients. We detect 170 asteroids with a signal-to-noise of at least $5$ in at least one of the ACT observing bands, which are centered near $90$, $150$, and $220$\,GHz. For each asteroid, we compare the ACT measured flux to predicted fluxes from the Near Earth Asteroid Thermal Model (NEATM) fit to {\it WISE} data. We confirm previous results that detected a deficit of flux at millimeter wavelengths. Moreover, we report a spectral characteristic to this deficit, such that the flux is relatively lower at $150$ and $220$\,GHz than at $90$\,GHz. Additionally, we find that the deficit in flux is greater for S-type asteroids than for C-type. 

\end{abstract}


\section{Introduction} \label{sec:intro}

The study of asteroids is critical for understanding the formation history of the solar system \citep[e.g.,][]{Michel2015} as they compose leftover material from the coalescence of the solar system. Asteroids have long been studied in the optical to infrared (IR) wavelengths, which encompass both the reflected and emitted peaks of asteroid emission. The Wide-field Infrared Survey Explorer ({\it WISE}) has detected hundreds of thousands of asteroids and measured their key features, including their sizes, emissivities at IR wavelengths, and temperatures \citep[e.g.,][]{Mainzer2011, Masiero2011}. These properties give clues about their material composition, and hence understanding them at multiple frequencies is key. Observations at the millimeter and sub-millimeter (mm and sub-mm) wavelengths supplement those made at IR frequencies \citep[e.g.,][]{Conklin1977, Johnston1982, Viikinkoski2015}. 
Emission in the mm and sub-mm is thermal in nature and originates from the depth of the attenuation length (i.e., several mm to cm) in the regolith, the unconsolidated surface of the asteroid. On the other hand, emission in the IR is primarily reflected light and originates much nearer the surface \citep[][]{Campbell1969}. 

Flux measurements of asteroids made in the sub-mm and mm have consistently found lower emission than expected from models fit to IR and optical data \citep{Johnston1982, Webster1988}. Historically, the reduced flux at these wavelengths has been interpreted as a drop in effective emissivity due to scattering of photons as they pass through the regolith \citep{Redman1992}. However, there is a growing body of evidence that the reduced flux actually arises from significantly lower temperatures than expected within the regolith compared to its surface \citep{Keihm2013}.

Progress on resolving this issue has been hindered by the lack of systematic surveys of asteroids in the mm and sub-mm. The field has long relied on targeted observations of asteroids, which require significant observatory resources \citep[see e.g.;][]{Muller2007, Chamberlain2007A}. Cosmic microwave background (CMB) experiments, which survey wide areas of the sky in the mm, offer the promise of large, well-calibrated catalogs of asteroids. Observations of asteroids with survey instruments have the advantage that they are `free' -- only an analysis is required to extract their fluxes from existing data, and no new observations are required. This is of great advantage in systematizing the study of asteroids in the mm and sub-mm.

Recently, the South Pole Telescope \citep[SPT;][]{Carlstrom2011} team reported flux measurements for a trio of main-belt asteroids \citep{Chichura2022}. For two of the three asteroids they detected, the flux measurements were consistent with predictions derived from {\it WISE} observations of unitary emissivity; for the third, the measured emissivity was $\epsilon = 0.64\pm 0.11$. SPT is located at the South Pole, and as such relatively few main-belt asteroids pass through its observing field, and those that do are only observable for a relatively short period of time. Observations from mid-latitude telescopes offer a better view of the ecliptic and correspondingly have more potential for asteroid observations.

In this paper, we present an analysis of over $100$ asteroids extracted from observations made by the Atacama Cosmology Telescope (ACT) between $2017$ and $2021$ from its location in Chile. We compare the fluxes measured by ACT to models calibrated with {\it WISE} data. We call the {\it WISE} model minus ACT data the ``model difference.'' We confirm that, in general, the model difference is negative, i.e., we observe a mm flux deficit. We study the dependence of the model difference on wavelength and asteroid class in a manner that has not previously been possible due to the relative scarcity of targeted observations. 

Data products from this paper, including normalized asteroid fluxes and phase curves, will be made available publicly. A companion paper is being prepared which will describe that data release and include instructions on how to utilize it.

This paper is structured as follows. In Section~\ref{sec:data}, we provide an overview of the ACT telescope, the {\it WISE} data set, and the data processing pipeline used to perform our analysis. In Section~\ref{sec:method}, we summarize the analysis used to investigate the model difference as well as to create light and phase curves. In Section~\ref{sec:res}, we present the results of that analysis. In Section~\ref{sec:disc}, we interpret these results with a particular view towards what they may mean for the regolith composition of asteroids. Finally, in Section~\ref{sec:conclusion}, we summarize and consider the opportunities presented by upcoming experiments to expand on this work.

\section{Data}
\label{sec:data}

\subsection{The Atacama Cosmology Telescope}
The Atacama Cosmology Telescope (ACT) was a $6$\,m off-axis Gregorian telescope located in the Atacama Desert in Chile \citep[][]{Fowler2007, Thornton2016} that was primarily used to make survey observations of the CMB from 2007 to 2022. ACT had three generations of receivers, most recently the Advanced ACTpol receiver \citep[AdvACT,][]{Henderson2016, Ho2016, Choi2020}. ACT observed the sky in six bands,  f030, f040, f090, f150, and f220, and f280 \citep{Li2021}. Of these, only the data from f090, f150, and f220 are used in this analysis. These bands are centered at approximately $90$, $150$, and $224$\,GHz, respectively, corresponding to diffraction-limited resolutions of $2.0\arcmin$, $1.4\arcmin$, and $1.0\arcmin$. 

\subsection{Depth-1 Maps}
\label{sec:depth-1}
ACT observed the sky by scanning back and forth at constant elevation, allowing the sky to pass through the observation track. Depth-1 maps are a single observation deep, in the sense that a given declination in the map only passes through the array of detectors once. They have the useful property that each pixel can be time-stamped with an accuracy of the time it takes a sky coordinate to drift through a detector array around 4 minutes. Since the asteroids do not move significantly on that timescale, we can accurately stack (Section~\ref{sec:stacking}) and phase-fold (Section~\ref{sec:phase_folding}) the asteroid maps. Depth-1 maps are made using the same maximum-likelihood framework as the normal ACT sky maps (see \cite{Dunner2013}, \cite{Naess2020}, and \cite{Aiola2020}), except that the conjugate gradient iteration used to invert the map-making equation is cut short, after 100 steps instead of 600, since the slower-converging large angular scales are irrelevant for the point-like objects that make up ACT's time-variable sky \citep[point sources generally converage after 10 steps, e.g.,][]{Marsden2014}. As with the normal maps, each frequency of each of ACT's dichroic detector arrays is mapped separately, resulting in a total of 29,175 depth-1 maps used in this work. All depth-1 maps were considered in this paper, although since not every depth-1 map contains an asteroid not all of them are actually used in the analysis. These depth-1 maps are part of the ACT data release 6 (DR6), and the exact depth-1 map-making procedure will be detailed in the upcoming DR6 paper. These maps will be of great use generally in searching for transient objects in the mm \citep[e.g.,][]{Li2023} as well as characterizing the variability of bright sources.

Two normalizations have been applied to the resultant fluxes, both of which are standard for ACT analyses. Firstly, there is a normalization of point sources fluxes to {\it Planck}, as outlined in \cite{Aiola2020}. While the f220 normalization is not yet public it has been computed in the same was as \cite{Aiola2020}. Secondly, we apply an effective bandcenter normalization which accounts for the varying effect of the bandpass with the spectral shape of the source being considered. These are also not yet publicly available for DR6, but will be made public with \cite{HasselfieldInPrep}. While these specific effecive bandcenters are not yet public, the methodology is a refinement of \cite{Swetz2011} and \cite{Thornton2016}, which are in turn based on \cite{Page2003}.

\subsection{{\it WISE}}
\label{sec:WISE_obs}
The  most up-to-date IR measurements of asteroid fluxes come from the {\it WISE} satellite \citep[][]{Wright2010}, and were analyzed by the Near Earth Object {\it WISE} (NEOWISE) group \citep{Mainzer2011}. {\it WISE} observed the whole sky in four IR bands centered at  $3.4$, $4.6$, $12$, and $22\,\mu$m, with $5\sigma$ point source sensitivities of  $0.08$, $0.11$, $1$, and $6$\,mJy, respectively. We use the preliminary results from \cite{Mainzer2011}; the IR sensitivity relative to the asteroid flux is much higher than the same in the mm, thus the preliminary release is sufficient for our purposes. These observations can be used to estimate the asteroid size, sub-solar temperature (which is the temperature at the hottest point of the asteroid), and emissivity. There are a number of models for asteroid fluxes \citep[see e.g.,][for an overview]{Mommert2018}, the most common of which is the Near Earth Asteroid Thermal Model \citep[NEATM;][]{Harris1998}. This model is used by the NEOWISE team and it can be used to generate predictions for the asteroid fluxes at the ACT frequencies. 


\begin{figure*}
    \centerline{
    \includegraphics[clip,trim=0.0cm 0.0cm 0.0cm 0.0cm,width=\textwidth]{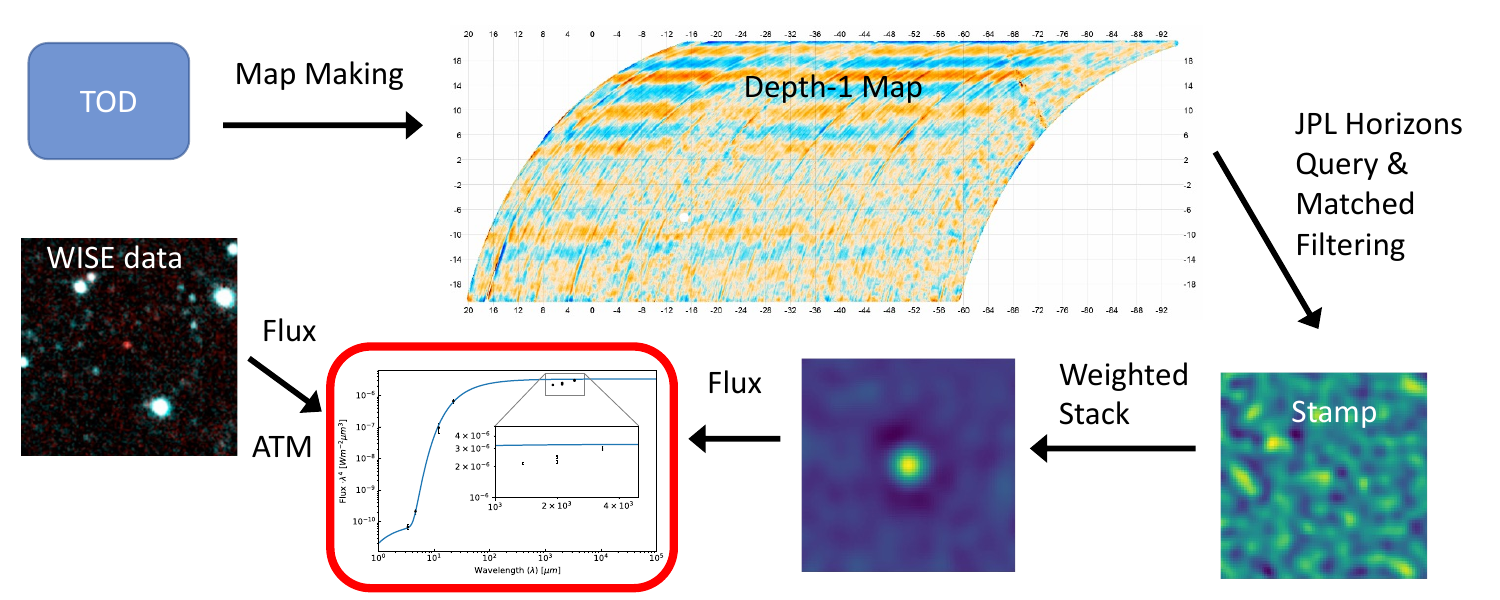}
    }
\caption{Summary of the data analysis pipeline used in this paper. Starting from the ACT time-ordered-data (TODs), we first construct depth-1 maps as described in Section~\ref{sec:depth-1}. We then use the JPL Horizons Query service provided by astroquery to obtain asteroid positions as a function of time, and then extract stamps from each depth-1 map at the location of each asteroid within it. For a given asteroid, we then stack the stamps, normalizing for the observational geometry and distance, as detailed in Section~\ref{sec:stacking}. Next, for each asteroid observed with high signal-to-noise by ACT, we use {\it WISE} IR observations of that asteroid, in combination with the ATM software package, to compute predictions from IR data for the flux at ACT's observing frequencies (Section~\ref{sec:ATM}). The differences between these predictions and the observed fluxes constitute the model difference. The model differences for (4) Vesta are shown in Figure~\ref{fig:atm_plot}.}
\label{fig:method_flow}
\end{figure*}


\section{Methods}
\label{sec:method}
To compute our measured flux, we first extract small maps centered on a given asteroid from the depth-1 maps, which we refer to as stamps. We then stack those stamps to obtain the asteroid flux. We also use {\it WISE} data in combination with the Asteroid Thermal Modeling software package \citep[ATM; ][]{Moeyens2020} to make predictions for the asteroid flux at the ACT observing frequencies. We then compare the observations to the predictions. The details of this workflow are given in Section~\ref{sec:stacking}, and a pictorial summary is shown in Figure~\ref{fig:method_flow}.
\subsection{ACT Stacking}
\label{sec:stacking}


\begin{figure*}
    \centerline{
    \includegraphics[clip,trim=0.0cm 0.0cm 0.0cm 0.0cm,width=\textwidth]{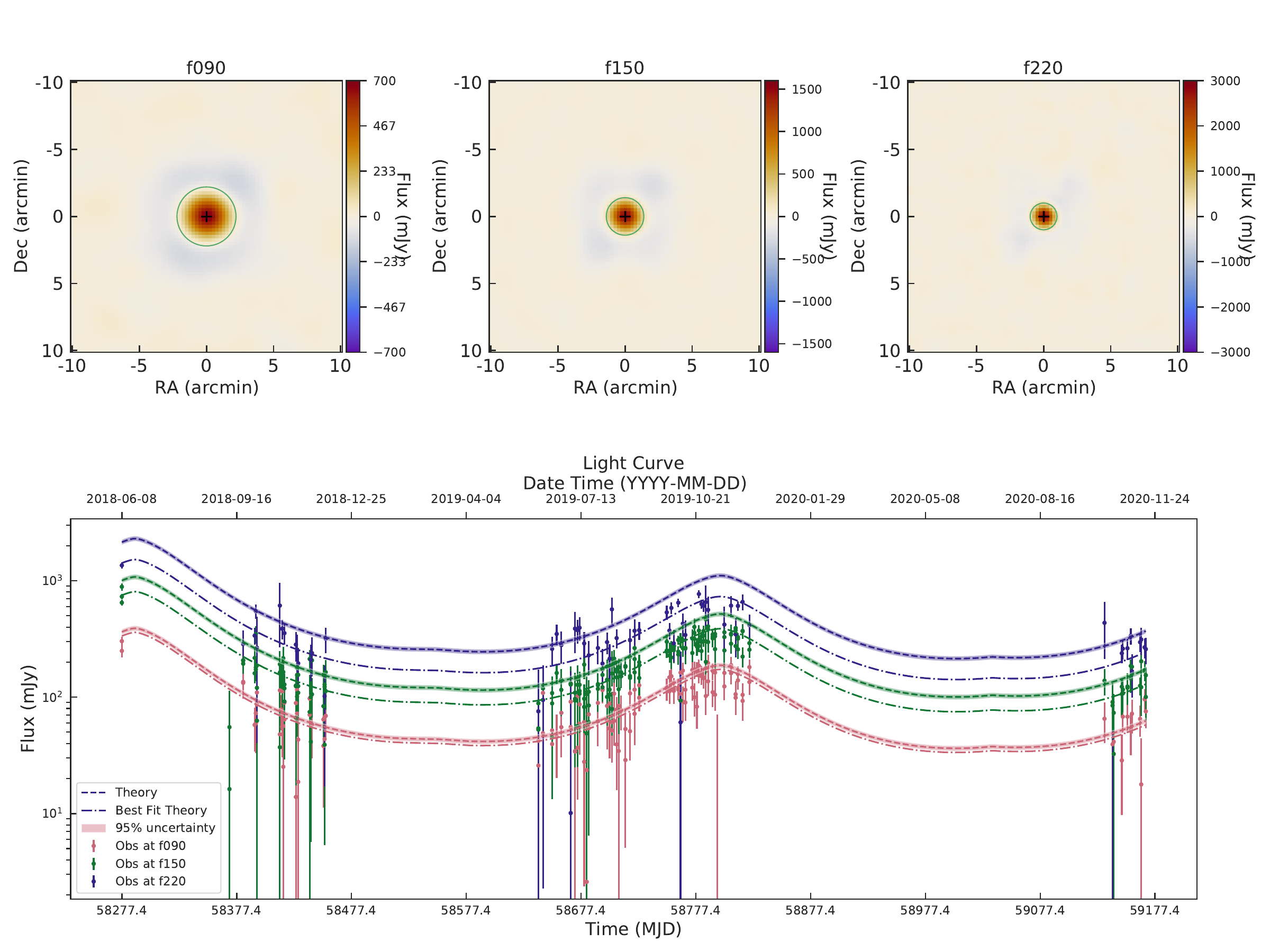}
    }
\caption{
        {\bf Top:} Stacked matched-filtered depth-1 maps of (4) Vesta showing the net flux at f090 (left), f150 (center), and f220 (right). The procedure for making these maps is described in Section~\ref{sec:stacking}. The map center is indicated by a black cross, while the beam full-width half-max at each frequency ($2.0$\arcmin, $1.4$\arcmin, and $1.0$\arcmin at f090, f150, and f220) is indicated by a green circle. Note the faint blue ring surrounding (4) Vesta at each frequency; this is ringing from the matched-filter used in the construction of the stamps (Section~\ref{sec:depth-1}). For this plot, we combined data across arrays using a pixel-by-pixel inverse-variance weight. This is {\it not} how we combined arrays in our analysis (Section~\ref{sec:stacking}), and is not an accurate method of capturing the noise properties of the maps. These images are for illustration only and were not used in scientific analysis. In our analysis, we do not combine maps across arrays. We instead only combined the central flux estimates in an inverse-variance way, including the $1\%$ bias discussed in Section~\ref{sec:stacking}.\\
        {\bf Bottom:} Light curve for (4) Vesta, generated as described in Section~\ref{sec:light_curve}. The three colored dashed lines with shading represent the average (with $95\%$ uncertainty) expected flux at each frequency based on {\it WISE} observations (Section~\ref{sec:ATM}), scaled according to Equation~\ref{eq:flux_weight}. The data lie below the {\it WISE} model line indicating that there is a deficit of mm flux. To show that the modeling scheme specified by  Equation~\ref{eq:max-like_f} is likely correct, we fitted the data at each frequency to the model curve times an overall amplitude, indicated by the dash-dotted line. The data points cluster around these best fit curves, indicating that our intrinsic relative flux modeling is likely correct. 
        This fit amplitude can be compared to the ratio of the amplitude of the stacked maps to the {\it WISE} expected fluxes as an internal consistency check. They are in excellent agreement for all frequencies.
}
\label{fig:vesta_sum}
\end{figure*}

We measure the flux of individual asteroids using a stacking method. For each asteroid, we consider each matched-filtered depth-1 map that contains the asteroid. We localize the asteroid within the map at the time of observation using the JPL Horizons service,\footnote{https://ssd.jpl.nasa.gov/horizons/} accessed using astroquery \citep{Ginsburg2019}. We take a $20'\times20'$ square stamp with $0.5'$ resolution centered at that position and tangent to the plane of the sky using the {\tt Pixell} software suite.\footnote{\url{https://github.com/simonsobs/pixell}} The asteroids are point sources in our maps;\footnote{The largest asteroids are $\sim 1000$\,km in diameter, with closest approach of $\sim1$\,AU, yielding an angular size $\lesssim 0.01'$, much lower than our resolution.} however, using larger stamps helps us filter out stamps with undesirable properties, for example those where the asteroid is near the edge of the map. To maximize the signal-to-noise (S/N), we apply a matched-filter to the depth-1 maps before cutting out and stacking the stamps. The filter removes both the large scales dominated by atmospheric noise and scales smaller than the beam. Given a sky map $m$ with noise covariance $N$ and beam matrix $B$, we form the matched-filter flux map $f$ as $f = \rho/\kappa$, with $\rho = B^T N^{-1}m$ and $\kappa = diag(B^T N^{-1} B)$. The associated flux uncertainty map is $1/\sqrt{\kappa}$. These are part of the standard ACT DR6 depth-1 release and are described in more detail in the upcoming ACT DR6 map paper. Our estimate for the asteroid's flux $F$ in a single depth-1 map is simply $f$ evaluated at the asteroid's location.

The observed flux of an asteroid varies from depth-1 map to depth-1 map due to the distance between the asteroid and earth/sun and also due to the changing observing angle. Removing this effect allows us to compare or combine measurements at different observing geometries; we call this removal ‘normalizing’. We normalize the observed flux by the expected flux in the Rayleigh-Jeans limit of the Standard Thermal Model (STM) \citep{Lebofsky1986}


\begin{align}
\label{eq:flux_weight}
 F_i &= F_0 \left(\frac{d_{\text{earth},i}}{1 \text{AU}}\right)^{-2}  \left(\frac{d_{\text{sun},i}}{1 \text{AU}}\right)^{-1/2} 10^{-0.004\alpha_i}\\
     &\equiv F_0 W_{i},    \notag
\end{align}
where $d_{\text{earth},i}$ and  $d_{\text{sun},i}$ are the Earth and Sun centered distance, and $\alpha_i$ is the Sun-asteroid-Earth phase angle in degrees, all at the time of the observation of stamp $i$. Here, $F_i$ is the observed asteroid flux and $F_0$ is the normalized asteroid flux, evaluated at $d_{\text{earth}} = d_{\text{sun}} = 1\text{AU}$. We use the STM to normalize the fluxes because it has a known closed form, and the geometrical scalings of the STM are the same as the NEATM. 
We then form the maximum-likelihood stacked flux estimate ($F_{stack,0}$) from the normalized stamps ($F_{stamp, i}$) as; 

\begin{align}
\label{eq:max-like_f}
    F_{stack, 0} &= (W^T N^{-1} W)^{-1} W^T N^{-1} F_{stamp, i}\\ \notag
        &= (\sum_i W_i N^{-1}_{ii} F_{stamp, i})/(\sum_i W_i^2 N^{-1}_{ii})
\end{align}
for $N$ the noise covariance matrix and $W$ the same as Equation~\ref{eq:flux_weight}.
Measuring the flux at the center of $F_{stack, 0}$ at the center pixel of the stack gives us a maximum likelihood estimate of the normalized flux. This flux can then be scaled by a particular Earth-asteroid distance, Sun-asteroid distance, and Sun-asteroid-Earth phase angle using Equation~\ref{eq:flux_weight} to compare with {\it WISE} observations (see Sec.~\ref{sec:WISE_obs}). Since the instrument beam is not yet well characterized for the daytime data, we only use nighttime data, specifically data from observations between 11pm and 11am coordinated universal time. 

As a consistency check for this method, we use the same pipeline to create flux maps for Uranus and compared them to dedicated scans of Uranus which are used for ACT calibration \citep{Hajian2011, Hasselfield2013}. Additionally, we computed the fractional difference in flux between arrays at the same frequency for all asteroids. We then combined this fractional difference in an inverse-variance weighted sense to get an average fraction difference between arrays at the same frequency. In both cases, the observed discrepancy is $<1\%$ between various arrays at the same frequency. This is consistent with the precision of the overall ACT calibration relative to {\it Planck}, which is of order $1\% $ \citep{Aiola2020} at f090 and f150, and $1.4\%$ at f220. This discrepancy is also consistent with dedicated scans of Uranus \citep{Hajian2011, Dunner2013, Hasselfield2013}. As a final consistency check, we compare the flux values from stamps of Uranus to estimates of its flux made directly from time-ordered-data (TODs); these agree within uncertainties.

In addition to the uncertainty in the calibration to {\it Planck}, there is a uncertainty in the beam size due to the effective frequency of observation; this effect is of order $1\%$ as well \citep[see][]{Marsden2014}. We combine these effects and add a $1.4\%$ systematic uncertainty term to the final flux at f090 and f150, and $1.8\%$ at f220. This systematic term is subdominant to the statistical one for all but four asteroids. 

A summary plot for the asteroid (4) Vesta is shown in Figure~\ref{fig:vesta_sum}. It includes stacked flux maps as described in this section, as well as a light curve (Section~\ref{sec:light_curve}).

\subsection{ATM Predictions}
\label{sec:ATM}
\begin{figure}
    \centerline{
    \includegraphics[width=\columnwidth]{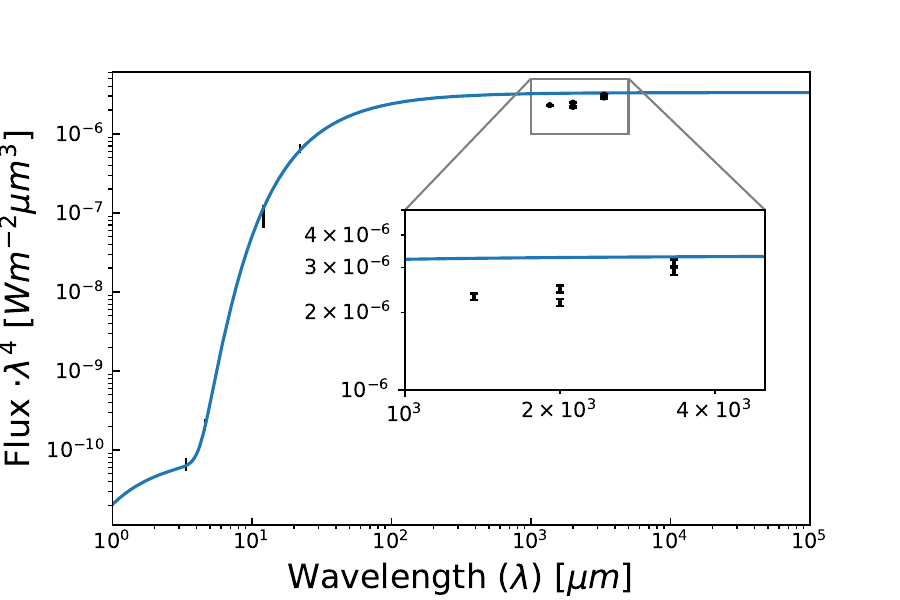}
}
\caption{Spectral energy distribution (SED) for the NEOWISE NEATM models as computed by ATM for the asteroid (4) Vesta, fit to observations in all four {\it WISE} bands (four left-most points). The SED has been scaled by the wavelength to the fourth power, so that blackbody regions are easy to distinguish as flat lines. The ACT data from stacking (Section~\ref{sec:stacking}) are the right-most points, with each array plotted separately. They are not included in the fit. The zoom shows the ACT deficit, i.e. that the ACT points lie bellow the blue line at all frequencies. The mm ACT fluxes are significantly deficient compared to the NEATM predictions ($6.8$ and $10.4\sigma$ at f150 and f220, respectively). Moreover, the deficiency is larger at shorter wavelengths. This pattern is consistent across our observations as detailed in Section~\ref{sec:res:stack_flux}.}
\label{fig:atm_plot}
\end{figure}

We use the ATM package, an open-source software which fits NEATM models to {\it WISE} data. The ATM package accurately reproduces the NEOWISE results, and also includes all the {\it WISE} data that are required to compare with ACT measured fluxes. The ATM package includes notebooks for fitting NEATM models to {\it WISE} data. Within these there are a number of prescriptions for treating the emissivities in the various {\it WISE} bands \citep[see][Chapter 3 for details of these prescriptions]{Moeyens2020}. We used the NEOWISE model, wherein the albedo is a free parameter, for comparison with ACT fluxes. When comparing our measured mm fluxes to the predictions from ATM, we scale the mm fluxes to the orbital configuration at the time of {\it WISE} observations using Equation~\ref{eq:flux_weight}. Each {\it WISE} observation of an asteroid typically includes $\sim 4-6$ individual exposures, which are spaced much less than a day apart. We evaluate Equation~\ref{eq:flux_weight} at the median time of each observation. The correction for the differing observation times is much less than $1\%$ and so it is not included in our analysis. An example comparison of ATM/{\it WISE} predictions and ACT observations is shown in Figure~\ref{fig:atm_plot}.

\subsection{Light Curves}
\label{sec:light_curve}
To generate the light curves, we use the stamps described at the beginning of Section~\ref{sec:stacking}. Instead of stacking them, we take the flux value from the center of each stamp and arrange them according to the time of observation of each stamp.
We also use fluxes from the NEATM model, scaling each model flux based on the Earth and Sun centered distance and Sun-asteroid-Earth phase angle according to Equation \ref{eq:flux_weight}. There is an associated  error for each flux in the stamps, which we use to generate the flux error bars. Figure \ref{fig:vesta_sum} provides an example light curve for (4) Vesta along with the NEATM scaled flux. The modulation in this light curve is apparent across all frequency bands and is a consequence of the change in observational geometry and distance. 

\subsection{Phase Curves}
\label{sec:phase_folding}


\begin{figure}
    \centering
    \includegraphics[width=\columnwidth, trim={2cm 8cm 1.9cm 7cm},clip]{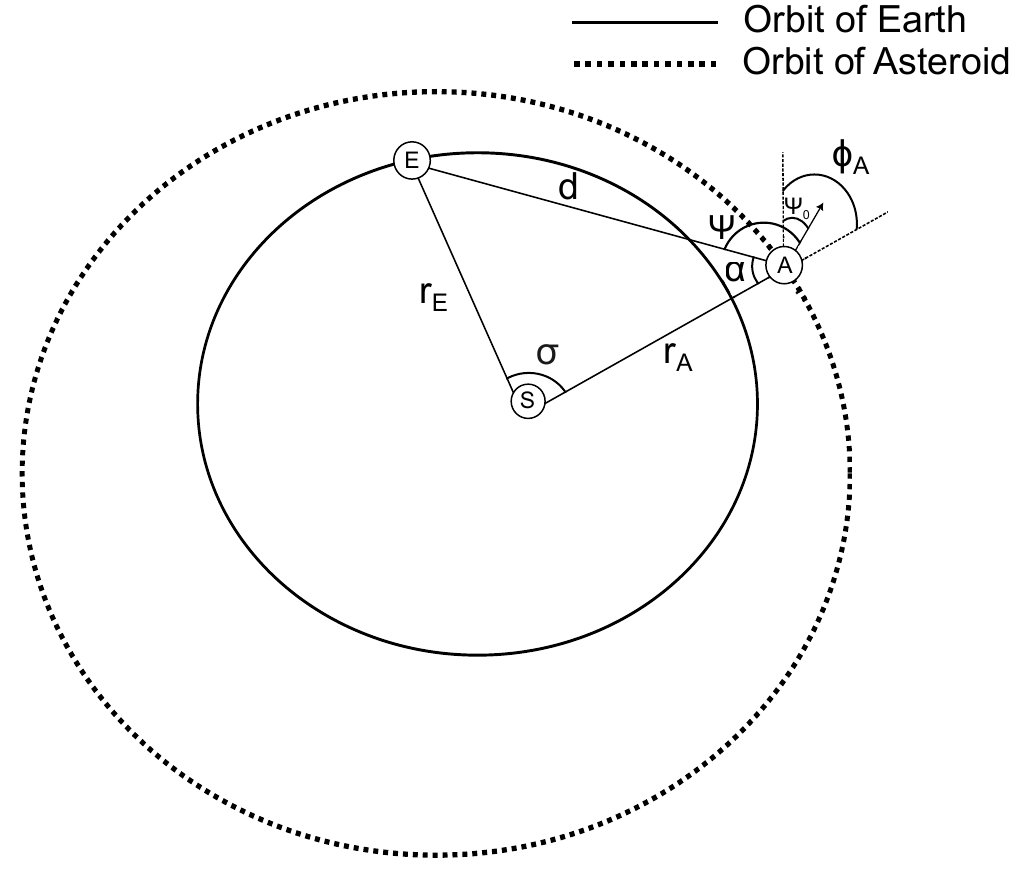}
    \caption{Schematic figure showing the angular relations between the Sun, Earth, and asteroid used to compute the asteroid phase. The faint vertical line represents the reference direction at $t_0$ at which time $\Psi_0 = 0$, which was chosen to be January 1, 1970. The Sun-Earth distance $r_e$ and Sun-Asteroid distance $r_a$, and Sun-Asteroid-Earth interior angle $\alpha$ are all computable from the ephemerides of the asteroid and Earth. Figure is not to scale.}
    \label{fig:phase_geometry}
\end{figure}


Phase curves are light curves that we fold in time by some frequency in order to detect periodic behavior in the asteroid flux. Specifically, we generate phase curves from the light curves in order to provide information about flux variations as a function of the asteroid sub-Earth longitude; that is, the line of longitude which intersects a line drawn from the center of the asteroid to the center of the Earth \citep[see e.g.,][]{Chamberlain2007A}. We refer to this as the phase of the asteroid. Figure~\ref{fig:phase_geometry} shows the relevant geometry. From the Figure, the Sun-Asteroid-Earth signed interior angle, $\alpha$, plus the difference between the observed phase, $\Psi_0$, and the rotational phase, $\Psi$, plus the asteroid longitude\footnote{Note this is not the ecliptic longitude, but the longitude relative to some reference time at which $\Psi_0 \equiv 0$. We have chosen this time to be midnight on January 1, 1970.} $\phi_A$ must be $\pi$. Rearranging, in radians we have;

\begin{equation}
    \label{eq:phase}
    \Psi = \Psi_0 - \alpha - \phi_A + \pi
\end{equation}
Computing $\Psi_0$ requires the rotational period of the asteroid, while computing $\alpha$ and $\phi_A$ requires the asteroid ephemerides, both of which we acquire from the Small-Body Database.\footnote{\url{https://ssd-api.jpl.nasa.gov/doc/sbdb.html}} There is one small correction to Equation~\ref{eq:phase}; the finite light travel time means that the phase at the time of observation is not the phase at the time the light was emitted from the asteroid. The Earth-Asteroid distance can change by several light-minutes, and this induces a non-negligible wobble in the phase. Since the light travel time is known, we subtract the light travel time from the time of observation. We then compute $\Psi_0$ modulo $2\pi$ and divide by $2\pi$ to obtain the dimensionless phase $p$.

In order to combine S/N across multiple arrays and frequencies, we fit the relative flux. For each observation $i$, we normalize the observed flux $F_i$ from Equation~\ref{eq:flux_weight} by multiplying by the appropriate factor to account for the observational distance and geometry (i.e. the weighted flux, Equation~\ref{eq:flux_weight}), then dividing by the average of the weighted flux at that frequency and array:

\begin{equation}
    \label{eq:norm_flux}
    F_{i, norm} = \frac{(F_i W_i)}{(\sum_N F_j W_j)/N}
\end{equation}



In addition to phase folding the light curves, we also plot the normalized flux and a best fit sinusoidal function for the phase curves. Our curve fit is a function of phase $p$ and is defined as follows:

\begin{equation}
    \label{eqn:best_fit_sin}
    F(p) = A_1\sin{(4\pi p + \phi_1)} +A_2\sin{(2\pi p + \phi_2)} + \delta \\
\end{equation}
where we fit for the amplitudes $A_1$ and $A_2$, phase factors $\phi_1$ and $\phi_2$, and offset $\delta$. Since we do not resolve the asteroids, we expect the primary modulation of the asteroid signal to be the observational cross section of the asteroid. Viewing the asteroid from one perspective and $180\deg$ from that perspective produces the same observational cross-section. Therefore, the phase curve frequency is twice the frequency of rotation; this is the $A_1$ term. We include the additional $A_2$ term to account for any potential modulation due to surface variations, such as low emissivity patches on the asteroid surface.

We use the MCMC implementation from \texttt{emcee} \citep{MacKey2013} to fit Equation~\ref{eqn:best_fit_sin} to the data with flat, uninformative priors on all parameters\footnote{Specifically, $|A_1|, |A_2|, |C| \leq 10$ and $-2\leq \phi_1, \phi_2 \leq2$}. We restricted our analysis to the 40 highest S/N asteroids. 


To determine the validity of the fits, we compared them to a pure constant fit using an F-test. An F-test can be used to compare nested models \citep{Allen1997}, where one model is a proper subset of the other. This allows us to compare the two sine fit to the constant fit. To perform the F-test, we construct the F-statistic:

\begin{equation}
    \label{eq:f_stat}
    F= \frac{\Delta \chi^2/\Delta \text{DoF}}{\chi_{nested}^2/\text{DoF}_{nested}}
\end{equation}
for $\Delta \chi^2$ and $\Delta \text{DoF}$ the change in $\chi^2$ and degrees of freedom between the nested and base model, and $\chi_{nested}^2$ and $\text{DoF}_{nested}$ the $\chi^2$ and degrees of freedom of the nested model. The F-statistic can then be converted into a p-value by using the F-distribution with appropriate number of degrees of freedom, which we then convert into a $\sigma$ value for convenience. Of our $40$ asteroids, we detect variation in the phase curve of the form given by Equation~\ref{eqn:best_fit_sin} at $>5\sigma$ for two of them, (6) Hebe and (15) Eunomia. Those fits are shown in Figure~\ref{fig:phase_results}.

Aside from variation of the form given by Equation~\ref{eqn:best_fit_sin}, we consider how well a given phase curve is described by a constant fit. To do so, we fit the phase curves to a constant value in the same manner as for the sine fits. Following \cite{Andrae2010}, we compare the resulting normalized residuals to a Gaussian distribution using the one-sided Kolmogorov-Smirnov (KS) test. The same test was performed with the normalized residuals of the binned data.

\section{Results}
\label{sec:res}

\subsection{Model Difference: Deficit}
\label{sec:res:stack_flux}





\begin{figure}
    \centerline{
    \includegraphics[width=\columnwidth]{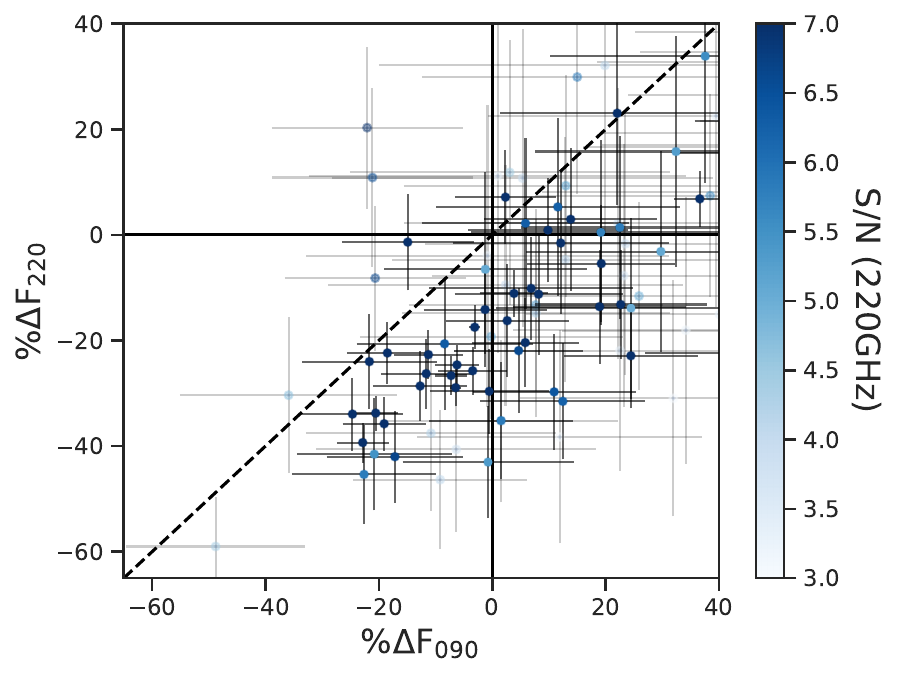}

}
\caption{ACT measured asteroid fluxes versus their {\it WISE} modeled counterparts for asteroids with S/N $>3$  at f220 and f090. Points to the left of $x=0$ have a deficit at f090 relative to the {\it WISE} predictions, and those that are below $y=0$ have a deficit at f220. Points with S/N $>5$ are in bold. The dashed line shows $\Delta F_{90} = \Delta F_{220}$ indicating an equal deficit at f090 and f220.88 out of 102 asteroids with S/N $>3$ at f220 and f090 have a larger relative deficit at f220 than at f090, equivalent to lying in the lower right half of this plot.}
\label{fig:deficit}
\end{figure}

We applied the method outlined in Section~\ref{sec:method} to $1200$ of the expected brightest asteroids. Of those 1200 asteroids, we detect 170 at $5\sigma$ in the stack in at least one band, meaning that $F_{\text{freq}} / \sigma_{\text{freq}} >5$ for one of the three frequency bands. Similarly, we detect 70 asteroids at $5\sigma$ in all bands. We detect 222 asteroids at $5\sigma$ when combining the detection significance across all three ACT bands. By this we mean combining the detection significance in each band in an inverse-variance manner. Using our entire catalog for which there are {\it WISE} data, we compared the measured flux at each frequency to the predictions from the model fit to {\it WISE} data (Section~\ref{sec:ATM}); this is the model difference. When the model difference is positive, we refer to it as an excess; when it is negative, a deficit. Here and throughout, the term larger is always used in an absolute sense, i.e., further from 0.
We found a consistent deficit at each frequency, with the deficit increasing with frequency. We considered the deficit for asteroids with WISE modeling with S/N $>5$ (177 asteroids) at the relevant frequency band, combining deficits via inverse-variance weighting. Defining the relative model difference to be:
\begin{equation}
    \label{eqn:delta_f}
    \Delta F_{90} \equiv (F_{90, ACT} - F_{90, {\it WISE}}) / F_{90, {\it WISE}}
\end{equation}
we find that the average relative model difference to be $(-4.0\pm 0.6\pm 1.4)\%$ at f090, $(-23.7\pm 0.4\pm 1.4)\%$ at f150, and $(-21.6\pm 0.9\pm 1.8)\%$ at f220, where the first uncertainty is $1\sigma$ statistical and the second is the $1.4$ or $1.8\%$ systematic gain uncertainty.\footnote{The deficit can be interpreted as an effective emissivity. However, due to the non-thermal distortion in the emission evidenced by the difference in deficit at various wavelengths, we do not do so in this paper.} This spectral dependence of the deficit is systematic in the sense that individual asteroids tend to have a larger deficit at f220 than at f090 (see Figure~\ref{fig:deficit}). For asteroids with S/N $>3$ in each ACT frequency band, $88$ out of $102$ have $\Delta F_{090} > \Delta F_{220}$, and$98$ out of $102$ have  $\Delta F_{090} > \Delta F_{150}$. Relatedly, $16$ such asteroids have $\Delta F_{090} > \Delta F_{220}$ at $2\sigma$ and $6$ have so at $3\sigma$. Only one asteroid with S/N $>3$ has $\Delta F_{090} < \Delta F_{150}$ or $\Delta F_{090} < \Delta F_{220}$ at the $1\sigma$ level, (63) Ausonia. Even in this case, only $\Delta F_{220} > \Delta F_{090}$ and not $\Delta F_{150}$.


\subsection{Model Difference: Excess}
\label{sec:res:excess}
Although the majority of the asteroids' model differences are $\leq 0$, a small number have statistically significant (S/N $> 3$) excess emission. Three asteroids ((511) Davida, (423) Diotima, and (611) Valeria) have excess emission at f090, and, moreover, (423) Diotima has $2.5\sigma$ excess at f150 and f220.

We inspected the light curves and individual stamps for each of these asteroids, as well as the ATM fits, to check for astrophysical interlopers or artifacts of filtering or map-making. In the case of (611) Valeria, the excess at f090 was found to be due to a single exposure where the asteroid was lying in the extended emission of a very bright point source. We instituted a check for this sort of scenario to all stamps, and the excess flux at f090 of (611) Valeria disappeared.

In the case of the other asteroids, we found nothing out of the ordinary. In the context of the full sample, that (511) Davida has $3\sigma$ excess flux in one frequency band is consistent with the statistical expectation. A $3\sigma$ outlier is about 1-in-300, and we have about $200$ asteroids with $3\sigma$ significant fluxes in any frequency band. (423) Diotima, on the other hand, is hard to accept as statistical coincidence. The light curves, individual stamps, and ATM fits for (423) Diotima all look normal, and the excess is statistically significant in each band; $3.6\sigma$ at f090, $2.9\sigma$ at f150 and $2.8\sigma$ at f220. Moreover, the fluxes follow the same behavior observed in other asteroids, wherein the flux at f090 ($130\pm 40\%$ excess) is higher than at f150 ($45\pm 16\%$) or f220 ($70 \pm 20\%$). Notably, (423) Diotima, a C-type asteroid (see Section~\ref{sec:disc:class}), has very low albedo \citep{Masiero2011}. 


\begin{table}
    \centering    
    \caption{Relative model difference for various asteroid classes. Note that in all bands the S-type asteroids have a larger deficit than the C-type asteroids. The M- and P-type asteroids also have generally large deficits while X-types are low deficit, although the low number of asteroids detected in each of these classes makes it difficult to say for certain. Note that both (511) Davida and (423) Diotima are C-type asteroids with high S/N; these two alone contribute significantly to the relatively low deficit at 90 GHz for this class of asteroids. The number of asteroids with total S/N $>5\sigma$ in each class is given in parentheses. The ``All'' row is all asteroids with total S/N $>5\sigma$ when combining across all bands, and includes the $1\%$ systematic error, as the statistical error is $\lesssim 1\%$.} 
    \begin{tabular}{cccc}
    \hline\hline\noalign{\smallskip}
        Class ($\#$)   & $\%\Delta \mathrm{F}_{090}$ & $\%\Delta \mathrm{F}_{150}$ & $\%\Delta \mathrm{F}_{220}$   \\
        \hline
        All (177) & $-4 \pm 2$ &  $-24 \pm 2$ &  $-22 \pm 3$\\
        C (63) & $0 \pm 3$ & $-20 \pm 2$ & $-23 \pm 3$  \\
        S (20) & $-19\pm 4$ & $-40 \pm 2$ & $-34 \pm 3$ \\
        M (4) & $-21\pm 22$ & $-22 \pm 10$ & $-15 \pm 14$ \\
        X (8) & $30\pm 21$ & $-11 \pm 10$ & $6 \pm 16$ \\
        P (9) & $-10\pm 9$ & $-32 \pm 4$ & $-24 \pm 6$ \\
        \hline

    \end{tabular}
    \label{tab:classes}
\end{table}


\subsection{Model Difference by Class}
\label{sec:res:class}

\begin{figure*}
    \centerline{
    \includegraphics[clip,trim=0.0cm 0.0cm 0.0cm 0.0cm,width=\textwidth, scale = 0.5]{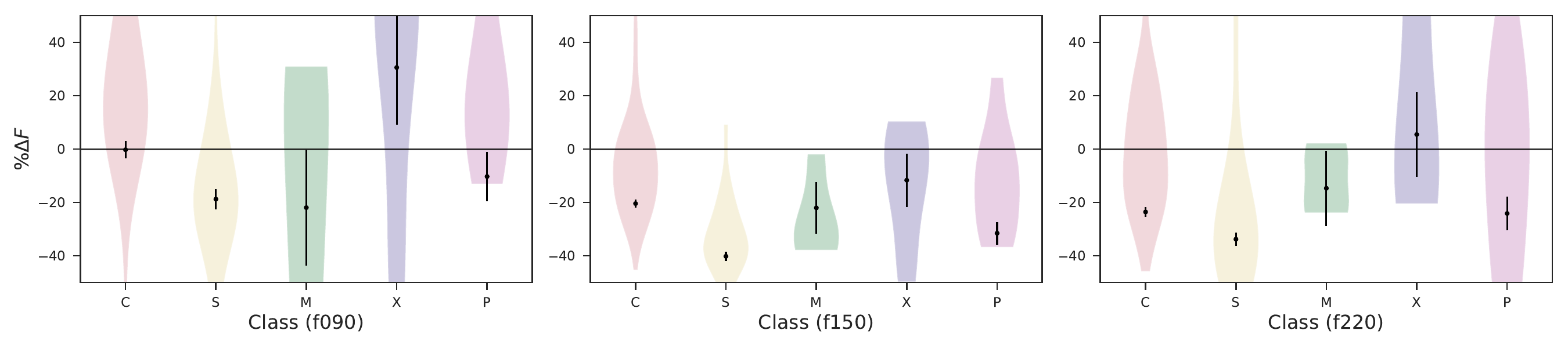}

    }
\caption{Plots showing the inverse-variance weighted average and population distribution of the model difference (Equation~\ref{eqn:delta_f}) at f090 (left), f150 (middle), and f220 (right) for various asteroid classes. The inverse-variance weighted mean and $1\sigma$ error on the mean are shown in black, while the colored violin-plots show the approximate unweighted population distribution. All asteroids in this plot have unambiguous types from the  Small-Body Database and have a total S/N $>5$ combining all frequencies. Note that the deficit for C-type asteroids is systematically smaller than for S-type asteroids, indicating that their composition is impacting the mm flux of those asteroids as compared to their IR flux. 
}
\label{fig:classes}
\end{figure*}

Under the assumption that the measured model differences are due to the composition of the regolith, we considered them as a function of asteroid class using the Tholen classification scheme \citep{Tholen1984}, which differentiates asteroids into $14$ types based on their spectra and albedo. By far the most common types of asteroids are C-types, which are dark and carbonaceous, and S-types, which tend to be more silicaceous. We used the Small-Body Database to obtain Tholen types for the asteroids detected by ACT. For the purposes of this section, we restricted our analysis to asteroids with unambiguous Tholen classifications, total S/N $>5$ across all bands, and {\it WISE} data. This results in $60$ C-type asteroids and $20$ S-type asteroids. 
We also detected $8$ X-type, $9$ P-type and $4$ M-type asteroids. The P- and M-type asteroids are defined by their optically reddish spectra \citep{Tholen1984}, 
and are distinguished from one another by their albedos, with P-types being of low albedo and M-types of moderate albedo. The X-type asteroids have the reddish spectra characteristic of M- and P-types, but do not have albedo measurements. 
For each class of asteroid, we computed the inverse-variance weighted average model difference at each of f090, f150, and f220. These are summarized in Table~\ref{tab:classes}. We also show plots of the distributions of model differences in Figure~\ref{fig:classes}.


\subsection{Phase Curves}
\label{res:phase_folding}





\begin{figure}
    \centerline{
    \includegraphics[width=\columnwidth, trim=1cm 0cm 0cm 0cm]{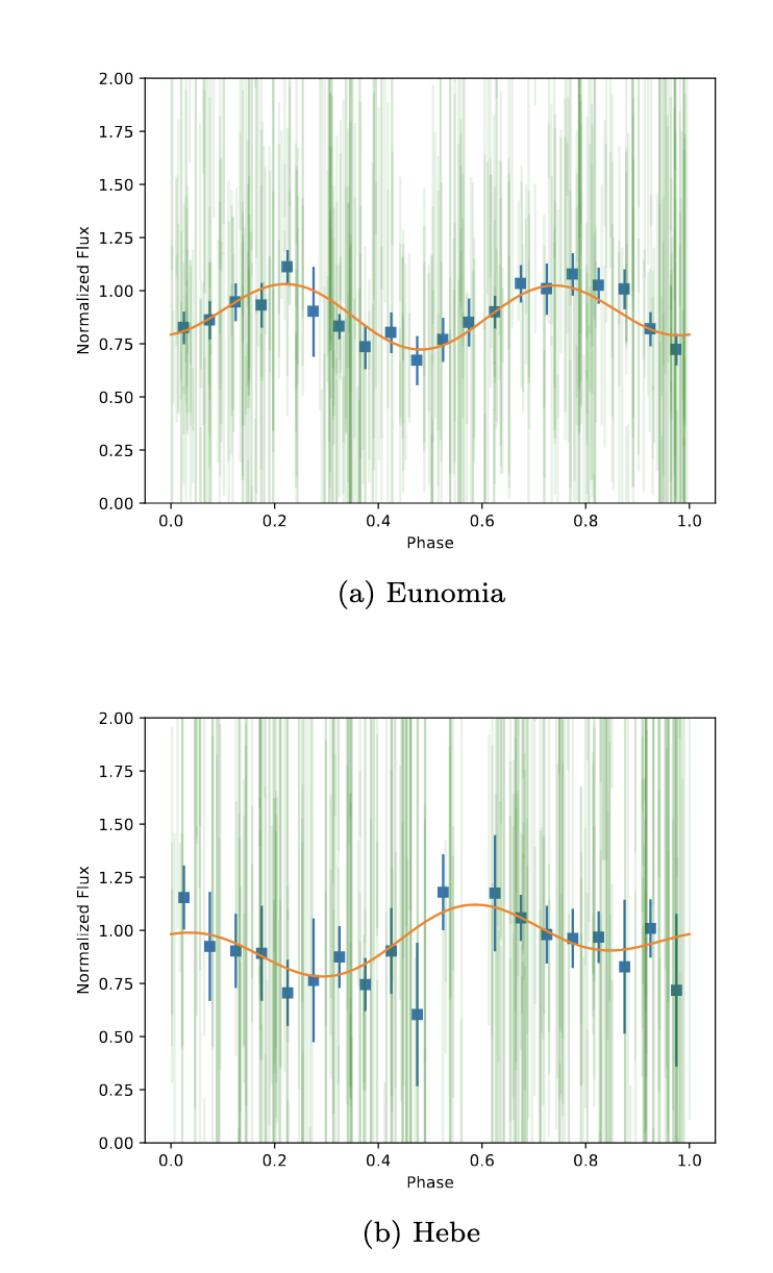}

}
\caption{Phase curves for (15) Eunomia and (6) Hebe, the two asteroids for which we make a tentative detection of phase variation. The orange line indicates the best fit of the two sine model from Equation~\ref{eqn:best_fit_sin}. The green bands indicate the individual observations, while the blue points are the inverse-variance weighted binning of those green points into 20 evenly spaced bins. The models are fit to the individual data points; the binned points have been plotted to guide the eye. One data binned data point for (6) Hebe has been omitted as it only contained one individual data point.} 
\label{fig:phase_results}
\end{figure}

We detect phase variation of the form given by Equation~\ref{eqn:best_fit_sin} for two asteroids at S/N greater than $5\sigma$ as determined by an F-test. Those asteroids are(15) Eunomia ($8.1\sigma$) and (6) Hebe ($7.5\sigma$). The best fit parameters for (15) Eunomia are $A_1 = 0.14^{+0.03}_{-0.03}$, $\phi_1 = 0.25^{+0.12}_{-0.17}$, $A_2 = 0.04^{+0.04}_{-0.06}$, $\phi_2 = 1.6^{+0.3}_{-1.3}$, and $\delta = 0.89^{+0.02}_{-0.02}$. For (6) Hebe, the best fit parameters are  $A_1 = 0.09^{+0.05}_{-0.06}$, $\phi_1 = 0.8^{+0.5}_{-0.3}$, $A_2 = 0.10^{+0.06}_{-0.05}$, $\phi_2 = 0.4^{+0.7}_{-0.3}$, and $\delta = 0.94^{+0.04}_{-0.04}$. Those models are shown in Figure~\ref{fig:phase_results}.

\begin{figure}
    \centerline{
    \includegraphics[width=\columnwidth, trim=1cm 0cm 0cm 0cm]{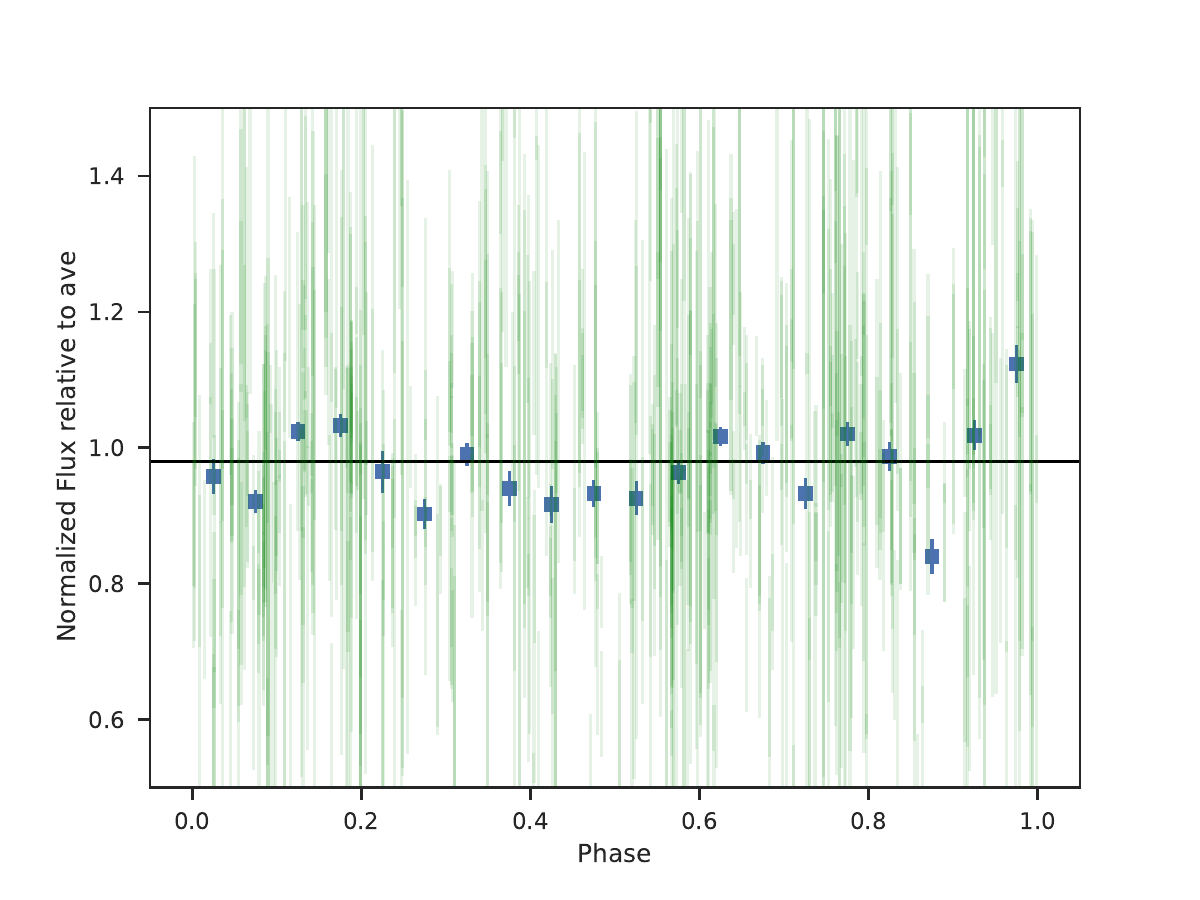}

}
\caption{Phase curve for (4) Vesta, one of the three asteroids whose phase curve is statistically inconsistent with constant flux and which is not well described by sinusoidal variation of the form given by Equation~\ref{eqn:best_fit_sin}. The green lines are the individual data points while the blue points are inverse-variance weighted bins of the green points. The black line is the best fit constant to the green data point. It is statistically inconsistent with the green data at$3.8\sigma$ and with the blue binned data at $5.3\sigma$}. 
\label{fig:vesta_phase}
\end{figure}

Of the asteroids for which we did not detect sinusoidal phase modulation, three showed statistically significant variation from constant flux. For (4) Vesta, we reject the null hypothesis of constant flux with phase at $5.3 (3.8)\sigma$ for the binned and unbinned data, for (7) Iris we reject it at $4.5 (3.5)\sigma$, and for (511) Davida we reject it at $3.3 (4.0)\sigma$. For these asteroids, the phase curves show variation in a more complicated manner than the simple sinusoidal variation given by Equation~\ref{eqn:best_fit_sin}. For reference, the phase curve for (4) Vesta is shown in Figure~\ref{fig:vesta_phase}.

\section{Discussion}
\label{sec:disc}
\subsection{Asteroid Fluxes}
On the whole, our observations of asteroids with ACT confirm previous findings of a flux deficit in the mm \citep{Johnston1982, Webster1988}. Our f090 deficit of $-4.0\pm0.6\%\pm 1.4\%$ is less than the deficits found by other works, which tend towards $-25\%$ \citep{Johnston1982, Webster1988}. On the other hand, our f150 ($-23.7\pm 0.4\%\pm 1.4\%$) and f220 deficits ($-21.6\pm 0.9\%\pm 1.8\%$) are in line with the literature. Since the deficits we quote are averaged across all asteroids, we do not necessarily expect them to exactly agree with observations of any one individual asteroid. Further, the decrease in flux with decreasing wavelength suggests that the material properties of the regolith change with depth, as has been suggested by e.g., \cite{Keihm2013}. However it is not possible to determine which material property, the emissivity or the temperature, changes without joint modeling of the ACT and {\it WISE} fluxes. We are currently working on extending the ATM software package to do so, and plan to present them in a future paper. Note that we observe the relative deficit to be decreasing with wavelength. Because longer wavelengths probe deeper into the regolith, this would require either the emissivity or the temperature to be increasing with depth. 
In general, one would instead expect the temperature to drop with increasing depth into the regolith; however,  without full joint modeling we cannot rule out emissivity variations as a source of the mm flux deficit. It may be that both variations in emissivity and temperature are involved in producing the deficit. Finally, it is somewhat counter-intuitive that the deficits are decreasing with wavelength; the shorter wavelength (f150/f220) observations do not probe the regolith as deeply as the f090 observations, and they are closer in frequency to the {\it WISE} observations to which the ATM was fit. On both counts one would expect this to make the short wavelength observations more similar to the IR, but the opposite is observed.

\subsection{Asteroid Flux Comparisons}
Given the large number of asteroids ACT detected, it is possible to directly compare our flux measurements with previous studies. 

\subsubsection{SPT}
We directly compare our observations of (324) Bamberga, (13) Egeria, (22) Kalliope to those made by SPT in \cite{Chichura2022}, specifically to the effective emissivity listed in Table 1 of that paper. We present the comparison in Table~\ref{tab:spt}. The ACT f090/f150 bands are slightly different than the SPT 95 and 150\,GHz channels. While this difference is small, we compare the effective emissivities, not the fluxes, which should be less sensitive to band differences. There resulting difference in the effective emissivities due to the differing band centers between ACT and SPT is relatively small, particularly in comparison to the uncertainties on the SPT results. In general these two works agree, with none of the asteroids showing a statistically significant difference in effective emissivity between the two studies. There is a tendency for the ACT measurements to have lower flux than the SPT measurements, but it is not statistically significant.


\begin{table*}
    \centering    
    \caption{Comparison of ACT effective emissitivy measurements to SPT effective emissivity measurements for the three asteroids significantly detected by SPT. The results are statistically consistent.}. 
    \begin{tabular}{ccccc}
    \hline\hline\noalign{\smallskip}
        Asteroid   & ACT f090 & SPT 95\,GHz & ACT f150 & SPT 150\,GHz   \\
        \hline
        (324) Bamberga & $1.0 \pm 0.2$ & $1.4 \pm 0.3$ & $0.80 \pm 0.10$ & $1.1\pm 0.1$ \\
        (13) Egeria & $0.79\pm 0.14$ & $<1.2$ & $0.71 \pm 0.06$ & $0.90 \pm 0.09$\\
        (22) Kalliope & $0.27\pm 0.27$ & $<0.47$ & $0.68\pm 0.12$ & $0.66\pm 0.11$\\
        \hline

    \end{tabular}
    \label{tab:spt}
\end{table*}


\subsubsection{Kitt Peak Antenna -- (1) Ceres}
\citet{Webster1988} used the Kitt Peak Antenna to make a number of observations of (1) Ceres at $89.5$ and $227$\,GHz \citep{Webster1988}. Due to the differences in observing frequency, we use the prescription for converting to brightness temperature\footnote{The brightness temperature is the temperature a blackbody would have to be to produce the observed flux at the observing frequency, and cannot be directly compared to the sub-solar temperatures that are estimated by the ATM fit.} used in \cite{Webster1988}, which is described in \cite{Johnston1982}. Our results, summarized in Table~\ref{tab:kp}, are in good agreement, given the large uncertainties on the \cite{Webster1988} results. Note that our f090 brightness temperature is significantly higher than that at f220; this is simply another way of stating that the relative deficit at f220 is higher than at f090.  


\begin{table}
    \centering    
    \caption{Comparison of ACT flux measurements to Kitt Peak observations of (1) Ceres. The brightness temperature is computed following  Equation 3 in \cite{Johnston1982}. We improve the uncertainty by about a factor of 2 at $90$\,GHz and about an order of magnitude at $220$\,GHz.} 
    \begin{tabular}{ccc}
    \hline\hline\noalign{\smallskip}
        Frequency (GHz)   & $\mathrm{T}_{\mathrm{B,n, ACT}}$ (K) & $\mathrm{T}_{\mathrm{B,n, KP}}$ (K)    \\
        \hline
        90 & $190\pm 10$ & $170\pm 20$ \\
        220 & $157.9\pm 1.5$ & $156\pm 25$ \\
        \hline

    \end{tabular}
    \label{tab:kp}
\end{table}



\subsubsection{ATCA -- (4) Vesta and (9) Metis}
\cite{Muller2007} made observations of the asteroids (4) Vesta and (9) Metis using the  Australia Telescope Compact Array (ATCA) at $93$ and $95.5$\,GHz. They included the time of observations, and so we were able to convert our flux measurements to their observing geometry and distance. We used the median time of observation to compute observing geometry and distance. The results change by $<0.5\%$ when using the two extremal times. Their observing frequencies are slightly different than ACT's, but from their work the difference in flux between $93$ and $95.5$\,GHz flux was $0.2$\,mJy, far less than the uncertainty on those measurements, so that comparisons to ACT are reasonable. These comparisons are given in Table~\ref{tab:atca}. We find a somewhat higher flux for (4) Vesta, by $\sim 2\sigma$. Moreover, we confirm their ``tentative'' result that the effective emissivity of (4) Vesta increases with wavelength. We find the effective emissivity to be $0.94\pm 0.01 $ at f090, $0.75 \pm 0.01$ at f150, and $0.70\pm 0.01 $ at f220, where we have combined the statistical uncertainty and $1\%$ systematic bias. Our effective emissivity is significantly higher than that found in \cite{Muller2007}, despite our measured fluxes being statistically consistent. This is due to differences in our modeled emission from (4) Vesta. \cite{Muller2007} undertook a much more detailed modeling of the expected thermal emission from (4) Vesta than we did, and so it should be considered the more accurate estimate. Note that our observed increase in relative flux with wavelength could also be explained by a rapidly rising temperature under the surface. \\

We also find higher emission at f090 for (9) Metis, although again this difference is only significant at $\sim 2\sigma$. Unfortunately we do not have {\it WISE} observations of (9) Metis, and so were unable to generate a NEATM model. As a rough comparison, we take the f150 fluxes and scale them to f090 and f220 using a simple $\nu^2$ scaling. This results in a expected flux of $28.1\pm 1.3$\,mJy at f090 as compared to a measured $40\pm 3$\,mJy, and an expected $168\pm 14$\,mJy as compared to a measured $150\pm 10$\,mJy at f220, suggesting that (9) Metis has a reddish spectrum. However, due to the lack of a NEATM model to compare to, this result should be considered tentative. 


\begin{table}
    \centering    
    \caption{Comparison of ACT flux measurements to ATCA observations of (4) Vesta and (9) Metis. We have used the median time of observation for ATCA to account for the observational geometry and distance. The units on all fluxes are mJy. In general the ACT fluxes are somewhat higher, although the difference is not statistically significant.}
    \begin{tabular}{cccc} 
    \hline\hline\noalign{\smallskip}
        Asteroid   & $\mathrm{F}_{\mathrm{ACT}} $ & $\mathrm{F}_{\mathrm{ATCA}} $  & $\mathrm{F}_{\mathrm{ATCA}} $   \\
            & 90\,GHz (mJy) & 93\,GHz (mJy) & 95.5\,GHz (mJy) \\
        
        \hline
        (4) Vesta & $172\pm 9$ & $147.3\pm 3.8$ & $147.5\pm 4.8$ \\
        (9) Metis & $40 \pm 3$ & $32.4\pm 1.1$ & $34.6\pm 1.2$ \\ 
        \hline

    \end{tabular}
    \label{tab:atca}
\end{table}


\subsection{Asteroid Fluxes by Class}
\label{sec:disc:class}
As shown in Table~\ref{tab:classes} and Figure~\ref{fig:classes}, the model difference (as computed in Section~\ref{sec:ATM}) is larger in S-type asteroids than in C-type ones. It is possible that this difference in deficits is caused by the different regolith properties of C-type (mostly carbon and ice) and S-type \citep[mostly silicacious;][]{Binzel1989, Bus2002} asteroids. 
Alternatively, it may be that the C-type asteroid spectra are better described as gray-bodies than those of S-type asteroids, so that the ATM modeling is more accurate for the C-type asteroids. However, there is good evidence that both types of asteroids are accurately modeled as gray-body \citep{Moeyens2020}. 
Finally, comparing the flux model differences at each frequency, there is evidence that the mm spectrum differs between C- and S-type asteroids (Section~\ref{sec:res:class}). For each asteroid, we compute

\begin{align*}
    \Delta \mathrm{F}_{090-150} \equiv  \%\Delta \mathrm{F}_{090} - \%\Delta \mathrm{F}_{150}\\
    \Delta \mathrm{F}_{150-220} \equiv \%\Delta \mathrm{F}_{150} - \%\Delta \mathrm{F}_{220}
\end{align*}

We then compute the average and standard deviations of these statistics over C- and S-type asteroids via bootstrapping. For C-type asteroids, we find $\Delta \mathrm{F}_{090-150} = 31\pm 5 \Delta\%$ \footnote{Note the statistic here is a difference of percentages and not a percentage change, i.e. it is percentage points.}, while for S-type asteroids,  $\Delta \mathrm{F}_{090-150} = 19 \pm 4\% \Delta \%$, somewhat inconsistent. On the other hand, the f150/f220 differences are consistent, with $\Delta \mathrm{F}_{150-220} = -9 \pm 3 \Delta \%$ for C-type asteroids and $\Delta \mathrm{F}_{150-220} = -9 \pm 6\Delta\%$ for S-type. A change in the spectral shape at mm wavelengths suggests at least a partially physical origin for the difference in C- versus S-type fluxes. A difference in the spectral shape could also potentially illuminate the physical mechanism, presumably in the regolith, sourcing that difference, and hence could indicate differences in regolith composition between C- and S-type asteroids. Observations of asteroids at higher frequencies, such as $280$ or $350$\,GHz, could shed further light on this. The upcoming Simons Observatory \citep[SO;][]{SO2019} and Fred Young Submillimeter Telescope \citep[FYST;][]{CCAT2023} will both provide that potential. 

Due to the low number of asteroids with high S/N in each class, we do not interpret the variation in model difference between M-, P- and X-type asteroids.


\begin{figure}
    \centerline{
    \includegraphics[width=\columnwidth]{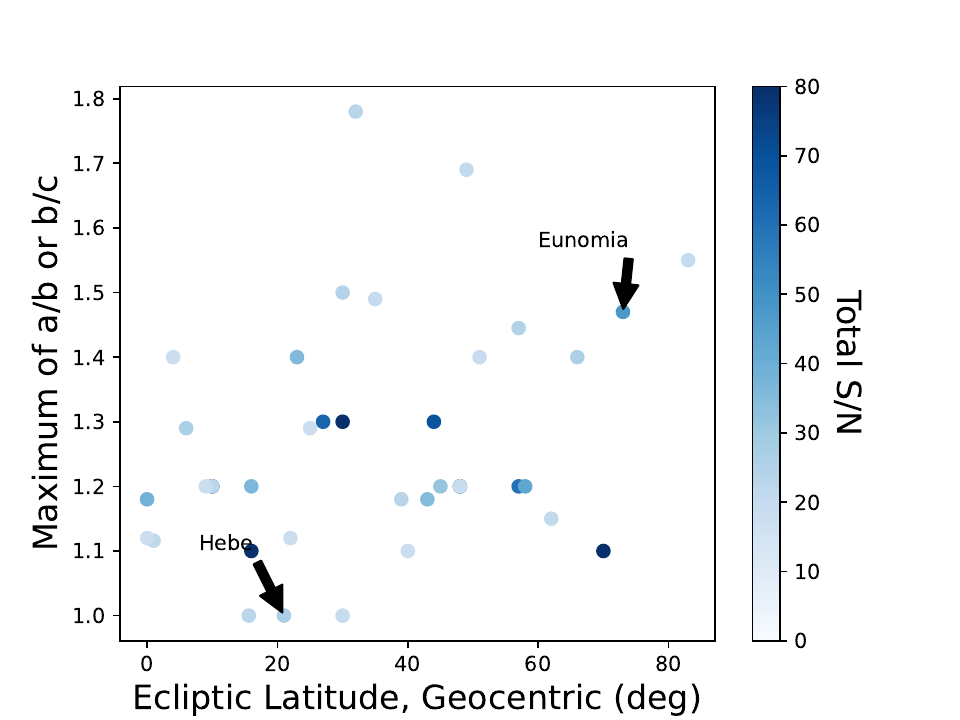}

}
\caption{Scatter plot of the ecliptic latitude of spin axis versus maximum ellipticity for the 40 highest S/N asteroids in our sample. For two asteroids ((554) Peraga and (375) Ursula) we were unable to find shape parameters, and hence they are omitted. The S/N scale has been restricted to a maximum of 80 to increase the dynamic contrast. (15) Eunomia and (6) Hebe have been labeled. (15) Eunomia is a noticeable outlier, while (6) Hebe lies nearer to the center of the population.} 
\label{fig:shape_spin}
\end{figure}

\subsection{Phase Curves}
\label{sec:disc:phase_curves}

(15) Eunomia and (6) Hebe have phase curves that are well described by sinusoid variation of the form of Equation~\ref{eqn:best_fit_sin}, with $\chi^2_{\text{red}}= 1.02$ and $0.99$, respectively. The best fit phase curve for (15) Eunomia indicates that the variation in emission is dominated by $A_1$, or half period term. Variation due to modulation of the asteroid cross-sectional area will lead to a large value for $A_1$, and indeed (15) Eunomia is quite non-spherical, with ellipticity parameters $a/b = 1.47, b/c = 1.0$ \citep{DeAngelis1995}, where $a$ is the longest semi-axis, $b$ the intermediate, and $c$ the shortest. On the other hand, for the more spherical asteroid (6) Hebe \citep[$a/b=1.1,b/c=1.2$,][]{Marsset2017}, we find $A_1\simeq A_2$, where $A_1$ and $A_2$ are the double and single frequency terms as defined in Equation~\ref{eqn:best_fit_sin}.

While we do detect phase variation for two asteroids in our sample, it is notable that we do not detect variation for other, higher S/N asteroids. While in general the S/N of phase curves increases with the S/N of the asteroid, the asteroid S/N is not the only determining factor. Firstly, as discussed above, the variation in asteroid cross-sectional area, and hence the asteroid shape, drives the $A_1$ variation term. This term is dominant in the phase curves of most asteroids \citep[although not all, see e.g.,][]{Chamberlain2007B}{}{}, so that variation is difficult to detect for highly spherical asteroids, even if they are high S/N. Secondly, the orientation of the rotational axis can suppress phase variation. As an extreme example, when the spin axis of an asteroid is pointed toward Earth, we observe no phase variation. In general, asteroids with highly inclined rotations will have their phase variation suppressed at some periods in their orbit. Finally, while sine functions do approximate the phase variation due to cross-sectional area, they do not necessarily describe all phase curves well. Surface features can easily cause other forms of variation, such as a step in emissivity, which would not be well described by a sine wave. Prior mm observations of (4) Vesta have shown highly irregular light curves, for example \citep{Muller2007}.

We performed a literature review for our 40 highest S/N asteroids, finding their shape parameters and spin-pole alignments. A scatter plot of these parameters, as well as our S/N, is shown  in Figure~\ref{fig:shape_spin}. (15) Eunomia is a noticeable outlier, with high S/N, high ellipticity, and a spin-axis that lies close to perpendicular to the ecliptic. (6) Hebe, on the other hand, is above average in S/N and spin-axis orientation, but more moderate in ellipticity. More extreme objects than (6) Hebe are not detected. Since we cannot cleanly explain our phase curve selection, we only tentatively claim detections for (15) Eunomia and (6) Hebe. 

Moreover, three asteroids have phase curves that are neither well described as constant nor sinusoidal. More in-depth analysis, perhaps in conjunction with existing observations such as \citet{Redman1992}, are needed to understand the origin of this variation.

\section{Conclusion}
\label{sec:conclusion}
We report flux measurements of asteroids in the millimeter using ACT. We detect $222$ asteroids at $5\sigma$ significance when combining significance across all three ACT bands, $170$ asteroids at $5\sigma$ significance in at least one band, and $70$ asteroids at $5\sigma$ significance in all bands. We confirm a deficit in mm flux as compared to expectations from IR measurements of those asteroids. Moreover, we detect a statistically significant spectral shape in the deficits, wherein the flux deficits are systematically larger at f150 and f220 than at f090. This suggests a more complicated source for the flux deficit than a simple change in effective emissivity with respect to the IR effective emissivity. However, we cannot determine the source conclusively without joint modeling of the IR and mm fluxes. 

Additionally, the relative mm flux as compared to {\it WISE} expectations is higher for C-type asteroids than for S-type. The spectrum of relative flux is also different between C- and S-type asteroids, with S-type asteroids having a flat spectrum between f150 and f220, while that of C-type asteroids falls in the same range. Both of these observations suggest compositional differences in the regoliths of C- versus S-type asteroids. Although {\it Planck} measured Zodiacal emission from families of asteroids \citep{Planck2014XIV}, this is the first systematic study of asteroid fluxes in the mm as a function of their class.

While we do not offer an explicit physical interpretation for these relative flux measurements, we do confirm the existence of a mm flux deficit, and we report a spectral and asteroid-class dependence of this deficit which suggests a physical origin in the regolith. 

We produce light curves for detected asteroids, as well as phase curves. For two asteroids, (15) Eunomia and (6) Hebe, we detect statistically significant variation in the phase curves of the sinusoidal form given by Equation~\ref{eqn:best_fit_sin}. However, we cannot adequately explain why phase variation in some asteroids is detected but variation in other asteroids is not. As such, caution should be exercised in ascribing any physical significance to that variation.

Looking to the future, the SO Large Aperture Telescope \citep[SO LAT;][]{Zhu2021, Parshley2018} will provide a significant increase in both sensitivity and frequency coverage over ACT, adding $280$\,GHz observations. The Fred Young Submillimeter Telescope (FYST) will complete a wide-field survey fully overlapping with the SO LAT survey area using its Prime-Cam \citep{CCAT2023} instrument to observe at $220$, $280$, $350$, $410$, and $850$\,GHz, adding to our ability to examine the spectra of asteroids in the sub-mm. Finally, looking to the far future, the proposed CMB-S4 \citep{S42016} and CMB-HD \citep{CMBHD2022} experiments would both represent over an order of magnitude improvement in sensitivity, and hence of asteroid detection. More sensitive observations over a wider wavelength range will further increase our ability to characterize the regolith composition of asteroids through their mm fluxes.

\section{Acknowledgments}
This work was supported by the U.S. National Science Foundation through awards AST-1440226, AST0965625 and AST-0408698 for the ACT project, as well as awards PHY-1214379 and PHY-0855887. The development of multichroic detectors and lenses was supported by NASA grants NNX13AE56G and NNX14AB58G. ACT operates in the Parque Astron\'{o}mico Atacama in northern Chile under the auspices of the Comisi\'{o}n Nacional de Investigaci\'{o}n Cient\'{i}fica y Tecnol\'{o}gica de Chile (CONICYT), now La Agencia Nacional de Investigaci\'{o}n y Desarrollo (ANID). Colleagues at AstroNorte and RadioSky provide logistical support and keep operations in Chile running smoothly.Computing for ACT was performed using the Princeton Research Computing resources at Princeton University, the National Energy Research Scientific Computing Center (NERSC), and the Niagara supercomputer at the SciNet HPC Consortium. SciNet is funded by the CFI under the auspices of Compute Canada, the Government of Ontario, the Ontario Research Fund–Research Excellence, and the University of Toronto.

John Orlowski-Scherer acknowledges support from the Trottier Space Institute Fellowship. This work was supported by a grant from the Simons Foundation (CCA 918271, PBL). Ricco C. Venterea acknowledges funding and support from the Nexus Scholars Program. Nick Battaglia acknowledges the support from NSF grant AST-1910021 and NASA grants 21-ADAP21-0114 and 21-ATP21-0129. Erminia Calabrese acknowledges support from the European Research Council (ERC) under the European Union’s Horizon 2020 research and innovation programme (Grant agreement No. 849169). Adam D. Hincks acknowledges support from the Sutton Family Chair in Science, Christianity and Cultures, from the Faculty of Arts and Science, University of Toronto, and from the Natural Sciences and Engineering Research Council of Canada (NSERC) [RGPIN-2023-05014, DGECR-2023-00180]. Yaqiong Li is supported by the KIC Postdoctoral Fellowship.  Crist\'obal Sif\'on acknowledges support from the Agencia Nacional de Investigaci\'on y Desarrollo (ANID) through FONDECYT grant no.\ 11191125 and BASAL project FB210003. Eve M. Vavagiakis acknowledges support from NSF award AST-2202237.

\bibliography{acteroids}{}

\begin{thebibliography}{}
\expandafter\ifx\csname natexlab\endcsname\relax\def\natexlab#1{#1}\fi
\providecommand{\url}[1]{\href{#1}{#1}}
\providecommand{\dodoi}[1]{doi:~\href{http://doi.org/#1}{\nolinkurl{#1}}}
\providecommand{\doeprint}[1]{\href{http://ascl.net/#1}{\nolinkurl{http://ascl.net/#1}}}
\providecommand{\doarXiv}[1]{\href{https://arxiv.org/abs/#1}{\nolinkurl{https://arxiv.org/abs/#1}}}

\bibitem[{{Abazajian} {et~al.}(2016){Abazajian}, {Adshead}, {Ahmed}, {Allen},
  {Alonso}, {Arnold}, {Baccigalupi}, {Bartlett}, {Battaglia}, {Benson},
  {Bischoff}, {Borrill}, {Buza}, {Calabrese}, {Caldwell}, {Carlstrom}, {Chang},
  {Crawford}, {Cyr-Racine}, {De Bernardis}, {de Haan}, {di Serego Alighieri},
  {Dunkley}, {Dvorkin}, {Errard}, {Fabbian}, {Feeney}, {Ferraro}, {Filippini},
  {Flauger}, {Fuller}, {Gluscevic}, {Green}, {Grin}, {Grohs}, {Henning},
  {Hill}, {Hlozek}, {Holder}, {Holzapfel}, {Hu}, {Huffenberger}, {Keskitalo},
  {Knox}, {Kosowsky}, {Kovac}, {Kovetz}, {Kuo}, {Kusaka}, {Le Jeune}, {Lee},
  {Lilley}, {Loverde}, {Madhavacheril}, {Mantz}, {Marsh}, {McMahon},
  {Meerburg}, {Meyers}, {Miller}, {Munoz}, {Nguyen}, {Niemack}, {Peloso},
  {Peloton}, {Pogosian}, {Pryke}, {Raveri}, {Reichardt}, {Rocha}, {Rotti},
  {Schaan}, {Schmittfull}, {Scott}, {Sehgal}, {Shandera}, {Sherwin}, {Smith},
  {Sorbo}, {Starkman}, {Story}, {van Engelen}, {Vieira}, {Watson}, {Whitehorn},
  \& {Kimmy Wu}}]{S42016}
{Abazajian}, K.~N., {Adshead}, P., {Ahmed}, Z., {et~al.} 2016, arXiv e-prints,
  arXiv:1610.02743, \dodoi{10.48550/arXiv.1610.02743}

\bibitem[{{Ade} {et~al.}(2019){Ade}, {Aguirre}, {Ahmed}, {Aiola}, {Ali},
  {Alonso}, {Alvarez}, {Arnold}, {Ashton}, {Austermann}, {Awan}, {Baccigalupi},
  {Baildon}, {Barron}, {Battaglia}, {Battye}, {Baxter}, {Bazarko}, {Beall},
  {Bean}, {Beck}, {Beckman}, {Beringue}, {Bianchini}, {Boada}, {Boettger},
  {Bond}, {Borrill}, {Brown}, {Bruno}, {Bryan}, {Calabrese}, {Calafut},
  {Calisse}, {Carron}, {Challinor}, {Chesmore}, {Chinone}, {Chluba}, {Cho},
  {Choi}, {Coppi}, {Cothard}, {Coughlin}, {Crichton}, {Crowley}, {Crowley},
  {Cukierman}, {D'Ewart}, {D{\"u}nner}, {de Haan}, {Devlin}, {Dicker},
  {Didier}, {Dobbs}, {Dober}, {Duell}, {Duff}, {Duivenvoorden}, {Dunkley},
  {Dusatko}, {Errard}, {Fabbian}, {Feeney}, {Ferraro}, {Flux{\`a}}, {Freese},
  {Frisch}, {Frolov}, {Fuller}, {Fuzia}, {Galitzki}, {Gallardo}, {Tomas Galvez
  Ghersi}, {Gao}, {Gawiser}, {Gerbino}, {Gluscevic}, {Goeckner-Wald}, {Golec},
  {Gordon}, {Gralla}, {Green}, {Grigorian}, {Groh}, {Groppi}, {Guan},
  {Gudmundsson}, {Han}, {Hargrave}, {Hasegawa}, {Hasselfield}, {Hattori},
  {Haynes}, {Hazumi}, {He}, {Healy}, {Henderson}, {Hervias-Caimapo}, {Hill},
  {Hill}, {Hilton}, {Hilton}, {Hincks}, {Hinshaw}, {Hlo{\v{z}}ek}, {Ho}, {Ho},
  {Howe}, {Huang}, {Hubmayr}, {Huffenberger}, {Hughes}, {Ijjas}, {Ikape},
  {Irwin}, {Jaffe}, {Jain}, {Jeong}, {Kaneko}, {Karpel}, {Katayama}, {Keating},
  {Kernasovskiy}, {Keskitalo}, {Kisner}, {Kiuchi}, {Klein}, {Knowles},
  {Koopman}, {Kosowsky}, {Krachmalnicoff}, {Kuenstner}, {Kuo}, {Kusaka},
  {Lashner}, {Lee}, {Lee}, {Leon}, {Leung}, {Lewis}, {Li}, {Li}, {Limon},
  {Linder}, {Lopez-Caraballo}, {Louis}, {Lowry}, {Lungu}, {Madhavacheril},
  {Mak}, {Maldonado}, {Mani}, {Mates}, {Matsuda}, {Maurin}, {Mauskopf}, {May},
  {McCallum}, {McKenney}, {McMahon}, {Meerburg}, {Meyers}, {Miller},
  {Mirmelstein}, {Moodley}, {Munchmeyer}, {Munson}, {Naess}, {Nati},
  {Navaroli}, {Newburgh}, {Nguyen}, {Niemack}, {Nishino}, {Orlowski-Scherer},
  {Page}, {Partridge}, {Peloton}, {Perrotta}, {Piccirillo}, {Pisano},
  {Poletti}, {Puddu}, {Puglisi}, {Raum}, {Reichardt}, {Remazeilles},
  {Rephaeli}, {Riechers}, {Rojas}, {Roy}, {Sadeh}, {Sakurai}, {Salatino},
  {Sathyanarayana Rao}, {Schaan}, {Schmittfull}, {Sehgal}, {Seibert}, {Seljak},
  {Sherwin}, {Shimon}, {Sierra}, {Sievers}, {Sikhosana}, {Silva-Feaver},
  {Simon}, {Sinclair}, {Siritanasak}, {Smith}, {Smith}, {Spergel}, {Staggs},
  {Stein}, {Stevens}, {Stompor}, {Suzuki}, {Tajima}, {Takakura}, {Teply},
  {Thomas}, {Thorne}, {Thornton}, {Trac}, {Tsai}, {Tucker}, {Ullom},
  {Vagnozzi}, {van Engelen}, {Van Lanen}, {Van Winkle}, {Vavagiakis},
  {Verg{\`e}s}, {Vissers}, {Wagoner}, {Walker}, {Ward}, {Westbrook},
  {Whitehorn}, {Williams}, {Williams}, {Wollack}, {Xu}, {Yu}, {Yu}, {Zago},
  {Zhang}, {Zhu}, \& {Simons Observatory Collaboration}}]{SO2019}
{Ade}, P., {Aguirre}, J., {Ahmed}, Z., {et~al.} 2019, \jcap, 2019, 056,
  \dodoi{10.1088/1475-7516/2019/02/056}

\bibitem[{{Aiola} {et~al.}(2020){Aiola}, {Calabrese}, {Maurin}, {Naess},
  {Schmitt}, {Abitbol}, {Addison}, {Ade}, {Alonso}, {Amiri}, {Amodeo},
  {Angile}, {Austermann}, {Baildon}, {Battaglia}, {Beall}, {Bean}, {Becker},
  {Bond}, {Bruno}, {Calafut}, {Campusano}, {Carrero}, {Chesmore}, {Cho},
  {Choi}, {Clark}, {Cothard}, {Crichton}, {Crowley}, {Darwish}, {Datta},
  {Denison}, {Devlin}, {Duell}, {Duff}, {Duivenvoorden}, {Dunkley},
  {D{\"u}nner}, {Essinger-Hileman}, {Fankhanel}, {Ferraro}, {Fox}, {Fuzia},
  {Gallardo}, {Gluscevic}, {Golec}, {Grace}, {Gralla}, {Guan}, {Hall},
  {Halpern}, {Han}, {Hargrave}, {Hasselfield}, {Helton}, {Henderson},
  {Hensley}, {Hill}, {Hilton}, {Hilton}, {Hincks}, {Hlo{\v{z}}ek}, {Ho},
  {Hubmayr}, {Huffenberger}, {Hughes}, {Infante}, {Irwin}, {Jackson}, {Klein},
  {Knowles}, {Koopman}, {Kosowsky}, {Lakey}, {Li}, {Li}, {Li}, {Lokken},
  {Louis}, {Lungu}, {MacInnis}, {Madhavacheril}, {Maldonado}, {Mallaby-Kay},
  {Marsden}, {McMahon}, {Menanteau}, {Moodley}, {Morton}, {Namikawa}, {Nati},
  {Newburgh}, {Nibarger}, {Nicola}, {Niemack}, {Nolta}, {Orlowski-Sherer},
  {Page}, {Pappas}, {Partridge}, {Phakathi}, {Pisano}, {Prince}, {Puddu}, {Qu},
  {Rivera}, {Robertson}, {Rojas}, {Salatino}, {Schaan}, {Schillaci}, {Sehgal},
  {Sherwin}, {Sierra}, {Sievers}, {Sifon}, {Sikhosana}, {Simon}, {Spergel},
  {Staggs}, {Stevens}, {Storer}, {Sunder}, {Switzer}, {Thorne}, {Thornton},
  {Trac}, {Treu}, {Tucker}, {Vale}, {Van Engelen}, {Van Lanen}, {Vavagiakis},
  {Wagoner}, {Wang}, {Ward}, {Wollack}, {Xu}, {Zago}, \& {Zhu}}]{Aiola2020}
{Aiola}, S., {Calabrese}, E., {Maurin}, L., {et~al.} 2020, \jcap, 2020, 047,
  \dodoi{10.1088/1475-7516/2020/12/047}

\bibitem[{Allen(1997)}]{Allen1997}
Allen, M.~P. 1997, Testing hypotheses in nested regression models (Boston, MA:
  Springer US), 113--117, \dodoi{10.1007/978-0-585-25657-3_24}

\bibitem[{{Andrae} {et~al.}(2010){Andrae}, {Schulze-Hartung}, \&
  {Melchior}}]{Andrae2010}
{Andrae}, R., {Schulze-Hartung}, T., \& {Melchior}, P. 2010, arXiv e-prints,
  arXiv:1012.3754, \dodoi{10.48550/arXiv.1012.3754}

\bibitem[{Binzel {et~al.}(1989)Binzel, Gehrels, \& Matthews}]{Binzel1989}
Binzel, R., Gehrels, T., \& Matthews, M. 1989, Asteroids II, Asteroids II No.
  v. 4 (University of Arizona Press)

\bibitem[{{Bus} \& {Binzel}(2002)}]{Bus2002}
{Bus}, S.~J., \& {Binzel}, R.~P. 2002, \icarus, 158, 146,
  \dodoi{10.1006/icar.2002.6856}

\bibitem[{{Campbell} \& {Ulrichs}(1969)}]{Campbell1969}
{Campbell}, M.~J., \& {Ulrichs}, J. 1969, \jgr, 74, 5867,
  \dodoi{10.1029/JB074i025p05867}

\bibitem[{{Carlstrom} {et~al.}(2011){Carlstrom}, {Ade}, {Aird}, {Benson},
  {Bleem}, {Busetti}, {Chang}, {Chauvin}, {Cho}, {Crawford}, {Crites}, {Dobbs},
  {Halverson}, {Heimsath}, {Holzapfel}, {Hrubes}, {Joy}, {Keisler}, {Lanting},
  {Lee}, {Leitch}, {Leong}, {Lu}, {Lueker}, {Luong-Van}, {McMahon}, {Mehl},
  {Meyer}, {Mohr}, {Montroy}, {Padin}, {Plagge}, {Pryke}, {Ruhl}, {Schaffer},
  {Schwan}, {Shirokoff}, {Spieler}, {Staniszewski}, {Stark}, {Tucker},
  {Vanderlinde}, {Vieira}, \& {Williamson}}]{Carlstrom2011}
{Carlstrom}, J.~E., {Ade}, P.~A.~R., {Aird}, K.~A., {et~al.} 2011, \pasp, 123,
  568, \dodoi{10.1086/659879}

\bibitem[{{CCAT-Prime Collaboration} {et~al.}(2023){CCAT-Prime Collaboration},
  {Aravena}, {Austermann}, {Basu}, {Battaglia}, {Beringue}, {Bertoldi},
  {Bigiel}, {Bond}, {Breysse}, {Broughton}, {Bustos}, {Chapman}, {Charmetant},
  {Choi}, {Chung}, {Clark}, {Cothard}, {Crites}, {Dev}, {Douglas}, {Duell},
  {D{\"u}nner}, {Ebina}, {Erler}, {Fich}, {Fissel}, {Foreman}, {Freundt},
  {Gallardo}, {Gao}, {Garc{\'\i}a}, {Giovanelli}, {Golec}, {Groppi}, {Haynes},
  {Henke}, {Hensley}, {Herter}, {Higgins}, {Hlo{\v{z}}ek}, {Huber}, {Huber},
  {Hubmayr}, {Jackson}, {Johnstone}, {Karoumpis}, {Keating}, {Komatsu}, {Li},
  {Magnelli}, {Matthews}, {Mauskopf}, {McMahon}, {Meerburg}, {Meyers},
  {Muralidhara}, {Murray}, {Niemack}, {Nikola}, {Okada}, {Puddu}, {Riechers},
  {Rosolowsky}, {Rossi}, {Rotermund}, {Roy}, {Sadavoy}, {Schaaf}, {Schilke},
  {Scott}, {Simon}, {Sinclair}, {Sivakoff}, {Stacey}, {Stutz}, {Stutzki},
  {Tahani}, {Thanjavur}, {Timmermann}, {Ullom}, {van Engelen}, {Vavagiakis},
  {Vissers}, {Wheeler}, {White}, {Zhu}, \& {Zou}}]{CCAT2023}
{CCAT-Prime Collaboration}, {Aravena}, M., {Austermann}, J.~E., {et~al.} 2023,
  \apjs, 264, 7, \dodoi{10.3847/1538-4365/ac9838}

\bibitem[{{Chamberlain} {et~al.}(2007{\natexlab{a}}){Chamberlain}, {Lovell}, \&
  {Sykes}}]{Chamberlain2007B}
{Chamberlain}, M.~A., {Lovell}, A.~J., \& {Sykes}, M.~V. 2007{\natexlab{a}},
  \icarus, 192, 448, \dodoi{10.1016/j.icarus.2007.08.003}

\bibitem[{{Chamberlain} {et~al.}(2007{\natexlab{b}}){Chamberlain}, {Sykes}, \&
  {Esquerdo}}]{Chamberlain2007A}
{Chamberlain}, M.~A., {Sykes}, M.~V., \& {Esquerdo}, G.~A. 2007{\natexlab{b}},
  \icarus, 188, 451, \dodoi{10.1016/j.icarus.2006.11.025}

\bibitem[{{Chichura} {et~al.}(2022){Chichura}, {Foster}, {Patel}, {Ossa-Jaen},
  {Ade}, {Ahmed}, {Anderson}, {Archipley}, {Austermann}, {Avva}, {Balkenhol},
  {Barry}, {Thakur}, {Beall}, {Benabed}, {Bender}, {Benson}, {Bianchini},
  {Bleem}, {Bouchet}, {Bryant}, {Byrum}, {Carlstrom}, {Carter}, {Cecil},
  {Chang}, {Chaubal}, {Chen}, {Chiang}, {Cho}, {Chou}, {Citron}, {Cliche},
  {Crawford}, {Crites}, {Cukierman}, {Daley}, {Denison}, {Dibert}, {Ding},
  {Dobbs}, {Dutcher}, {Everett}, {Feng}, {Ferguson}, {Fu}, {Galli},
  {Gallicchio}, {Gambrel}, {Gardner}, {George}, {Goeckner-Wald}, {Gualtieri},
  {Guns}, {Gupta}, {Guyser}, {de Haan}, {Halverson}, {Harke-Hosemann},
  {Harrington}, {Henning}, {Hilton}, {Hivon}, {Holder}, {Holzapfel}, {Hood},
  {Howe}, {Hrubes}, {Huang}, {Hubmayr}, {Irwin}, {Jeong}, {Jonas}, {Jones},
  {Khaire}, {Knox}, {Kofman}, {Korman}, {Kubik}, {Kuhlmann}, {Kuo}, {Lee},
  {Leitch}, {Li}, {Lowitz}, {Lu}, {Marrone}, {McMahon}, {Meyer}, {Michalik},
  {Millea}, {Mocanu}, {Montgomery}, {Moran}, {Nadolski}, {Natoli}, {Nguyen},
  {Nibarger}, {Noble}, {Novosad}, {Omori}, {Padin}, {Pan}, {Paschos}, {Patil},
  {Pearson}, {Phadke}, {Posada}, {Prabhu}, {Pryke}, {Quan}, {Rahlin},
  {Reichardt}, {Riebel}, {Riedel}, {Rouble}, {Ruhl}, {Saliwanchik}, {Sayre},
  {Schaffer}, {Schiappucci}, {Shirokoff}, {Sievers}, {Smecher}, {Sobrin},
  {Springmann}, {Stark}, {Stephen}, {Story}, {Suzuki}, {Tandoi}, {Thompson},
  {Thorne}, {Tucker}, {Umilta}, {Vale}, {Veach}, {Vieira}, {Wang}, {Whitehorn},
  {Wu}, {Yefremenko}, {Yoon}, \& {Young}}]{Chichura2022}
{Chichura}, P.~M., {Foster}, A., {Patel}, C., {et~al.} 2022, \apj, 936, 173,
  \dodoi{10.3847/1538-4357/ac89ec}

\bibitem[{{Choi} {et~al.}(2020){Choi}, {Hasselfield}, {Ho}, {Koopman}, {Lungu},
  {Abitbol}, {Addison}, {Ade}, {Aiola}, {Alonso}, {Amiri}, {Amodeo}, {Angile},
  {Austermann}, {Baildon}, {Battaglia}, {Beall}, {Bean}, {Becker}, {Bond},
  {Bruno}, {Calabrese}, {Calafut}, {Campusano}, {Carrero}, {Chesmore}, {Cho},
  {Clark}, {Cothard}, {Crichton}, {Crowley}, {Darwish}, {Datta}, {Denison},
  {Devlin}, {Duell}, {Duff}, {Duivenvoorden}, {Dunkley}, {D{\"u}nner},
  {Essinger-Hileman}, {Fankhanel}, {Ferraro}, {Fox}, {Fuzia}, {Gallardo},
  {Gluscevic}, {Golec}, {Grace}, {Gralla}, {Guan}, {Hall}, {Halpern}, {Han},
  {Hargrave}, {Henderson}, {Hensley}, {Hill}, {Hilton}, {Hilton}, {Hincks},
  {Hlo{\v{z}}ek}, {Hubmayr}, {Huffenberger}, {Hughes}, {Infante}, {Irwin},
  {Jackson}, {Klein}, {Knowles}, {Kosowsky}, {Lakey}, {Li}, {Li}, {Li},
  {Lokken}, {Louis}, {MacInnis}, {Madhavacheril}, {Maldonado}, {Mallaby-Kay},
  {Marsden}, {Maurin}, {McMahon}, {Menanteau}, {Moodley}, {Morton}, {Naess},
  {Namikawa}, {Nati}, {Newburgh}, {Nibarger}, {Nicola}, {Niemack}, {Nolta},
  {Orlowski-Sherer}, {Page}, {Pappas}, {Partridge}, {Phakathi}, {Prince},
  {Puddu}, {Qu}, {Rivera}, {Robertson}, {Rojas}, {Salatino}, {Schaan},
  {Schillaci}, {Schmitt}, {Sehgal}, {Sherwin}, {Sierra}, {Sievers}, {Sifon},
  {Sikhosana}, {Simon}, {Spergel}, {Staggs}, {Stevens}, {Storer}, {Sunder},
  {Switzer}, {Thorne}, {Thornton}, {Trac}, {Treu}, {Tucker}, {Vale}, {Van
  Engelen}, {Van Lanen}, {Vavagiakis}, {Wagoner}, {Wang}, {Ward}, {Wollack},
  {Xu}, {Zago}, \& {Zhu}}]{Choi2020}
{Choi}, S.~K., {Hasselfield}, M., {Ho}, S.-P.~P., {et~al.} 2020, \jcap, 2020,
  045, \dodoi{10.1088/1475-7516/2020/12/045}

\bibitem[{{Conklin} {et~al.}(1977){Conklin}, {Ulich}, {Dickel}, \&
  {Ther}}]{Conklin1977}
{Conklin}, E.~K., {Ulich}, B.~L., {Dickel}, J.~R., \& {Ther}, D.~T. 1977, in
  IAU Colloq. 39: Comets, Asteroids, Meteorites: Interrelations, Evolution and
  Origins, ed. A.~H. {Delsemme}, 257--261

\bibitem[{{De Angelis}(1995)}]{DeAngelis1995}
{De Angelis}, G. 1995, \planss, 43, 649, \dodoi{10.1016/0032-0633(94)00151-G}

\bibitem[{{D{\"u}nner} {et~al.}(2013){D{\"u}nner}, {Hasselfield}, {Marriage},
  {Sievers}, {Acquaviva}, {Addison}, {Ade}, {Aguirre}, {Amiri}, {Appel},
  {Barrientos}, {Battistelli}, {Bond}, {Brown}, {Burger}, {Calabrese},
  {Chervenak}, {Das}, {Devlin}, {Dicker}, {Bertrand Doriese}, {Dunkley},
  {Essinger-Hileman}, {Fisher}, {Gralla}, {Fowler}, {Hajian}, {Halpern},
  {Hern{\'a}ndez-Monteagudo}, {Hilton}, {Hilton}, {Hincks}, {Hlozek},
  {Huffenberger}, {Hughes}, {Hughes}, {Infante}, {Irwin}, {Baptiste Juin},
  {Kaul}, {Klein}, {Kosowsky}, {Lau}, {Limon}, {Lin}, {Louis}, {Lupton},
  {Marsden}, {Martocci}, {Mauskopf}, {Menanteau}, {Moodley}, {Moseley},
  {Netterfield}, {Niemack}, {Nolta}, {Page}, {Parker}, {Partridge}, {Quintana},
  {Reid}, {Sehgal}, {Sherwin}, {Spergel}, {Staggs}, {Swetz}, {Switzer},
  {Thornton}, {Trac}, {Tucker}, {Warne}, {Wilson}, {Wollack}, \&
  {Zhao}}]{Dunner2013}
{D{\"u}nner}, R., {Hasselfield}, M., {Marriage}, T.~A., {et~al.} 2013, \apj,
  762, 10, \dodoi{10.1088/0004-637X/762/1/10}

\bibitem[{{Foreman-Mackey} {et~al.}(2013){Foreman-Mackey}, {Hogg}, {Lang}, \&
  {Goodman}}]{MacKey2013}
{Foreman-Mackey}, D., {Hogg}, D.~W., {Lang}, D., \& {Goodman}, J. 2013, \pasp,
  125, 306, \dodoi{10.1086/670067}

\bibitem[{{Fowler} {et~al.}(2007){Fowler}, {Niemack}, {Dicker}, {Aboobaker},
  {Ade}, {Battistelli}, {Devlin}, {Fisher}, {Halpern}, {Hargrave}, {Hincks},
  {Kaul}, {Klein}, {Lau}, {Limon}, {Marriage}, {Mauskopf}, {Page}, {Staggs},
  {Swetz}, {Switzer}, {Thornton}, \& {Tucker}}]{Fowler2007}
{Fowler}, J.~W., {Niemack}, M.~D., {Dicker}, S.~R., {et~al.} 2007, \ao, 46,
  3444, \dodoi{10.1364/AO.46.003444}

\bibitem[{{Ginsburg} {et~al.}(2019){Ginsburg}, {Sip{\H o}cz}, {Brasseur},
  {Cowperthwaite}, {Craig}, {Deil}, {Guillochon}, {Guzman}, {Liedtke}, {Lian
  Lim}, {Lockhart}, {Mommert}, {Morris}, {Norman}, {Parikh}, {Persson},
  {Robitaille}, {Segovia}, {Singer}, {Tollerud}, {de Val-Borro}, {Valtchanov},
  {Woillez}, {The Astroquery collaboration}, \& {a subset of the astropy
  collaboration}}]{Ginsburg2019}
{Ginsburg}, A., {Sip{\H o}cz}, B.~M., {Brasseur}, C.~E., {et~al.} 2019, \aj,
  157, 98, \dodoi{10.3847/1538-3881/aafc33}

\bibitem[{{Hajian} {et~al.}(2011){Hajian}, {Acquaviva}, {Ade}, {Aguirre},
  {Amiri}, {Appel}, {Barrientos}, {Battistelli}, {Bond}, {Brown}, {Burger},
  {Chervenak}, {Das}, {Devlin}, {Dicker}, {Bertrand Doriese}, {Dunkley},
  {D{\"u}nner}, {Essinger-Hileman}, {Fisher}, {Fowler}, {Halpern},
  {Hasselfield}, {Hern{\'a}ndez-Monteagudo}, {Hilton}, {Hilton}, {Hincks},
  {Hlozek}, {Huffenberger}, {Hughes}, {Hughes}, {Infante}, {Irwin}, {Baptiste
  Juin}, {Kaul}, {Klein}, {Kosowsky}, {Lau}, {Limon}, {Lin}, {Lupton},
  {Marriage}, {Marsden}, {Mauskopf}, {Menanteau}, {Moodley}, {Moseley},
  {Netterfield}, {Niemack}, {Nolta}, {Page}, {Parker}, {Partridge}, {Reid},
  {Sehgal}, {Sherwin}, {Sievers}, {Spergel}, {Staggs}, {Swetz}, {Switzer},
  {Thornton}, {Trac}, {Tucker}, {Warne}, {Wollack}, \& {Zhao}}]{Hajian2011}
{Hajian}, A., {Acquaviva}, V., {Ade}, P. A.~R., {et~al.} 2011, \apj, 740, 86,
  \dodoi{10.1088/0004-637X/740/2/86}

\bibitem[{{Harris}(1998)}]{Harris1998}
{Harris}, A.~W. 1998, \icarus, 131, 291, \dodoi{10.1006/icar.1997.5865}

\bibitem[{{Hasselfield}(in prep.)}]{HasselfieldInPrep}
{Hasselfield}, M. in prep.

\bibitem[{{Hasselfield} {et~al.}(2013){Hasselfield}, {Moodley}, {Bond}, {Das},
  {Devlin}, {Dunkley}, {D{\"u}nner}, {Fowler}, {Gallardo}, {Gralla}, {Hajian},
  {Halpern}, {Hincks}, {Marriage}, {Marsden}, {Niemack}, {Nolta}, {Page},
  {Partridge}, {Schmitt}, {Sehgal}, {Sievers}, {Staggs}, {Swetz}, {Switzer}, \&
  {Wollack}}]{Hasselfield2013}
{Hasselfield}, M., {Moodley}, K., {Bond}, J.~R., {et~al.} 2013, \apjs, 209, 17,
  \dodoi{10.1088/0067-0049/209/1/17}

\bibitem[{{Henderson} {et~al.}(2016){Henderson}, {Allison}, {Austermann},
  {Baildon}, {Battaglia}, {Beall}, {Becker}, {De Bernardis}, {Bond},
  {Calabrese}, {Choi}, {Coughlin}, {Crowley}, {Datta}, {Devlin}, {Duff},
  {Dunkley}, {D{\"u}nner}, {van Engelen}, {Gallardo}, {Grace}, {Hasselfield},
  {Hills}, {Hilton}, {Hincks}, {Hloẑek}, {Ho}, {Hubmayr}, {Huffenberger},
  {Hughes}, {Irwin}, {Koopman}, {Kosowsky}, {Li}, {McMahon}, {Munson}, {Nati},
  {Newburgh}, {Niemack}, {Niraula}, {Page}, {Pappas}, {Salatino}, {Schillaci},
  {Schmitt}, {Sehgal}, {Sherwin}, {Sievers}, {Simon}, {Spergel}, {Staggs},
  {Stevens}, {Thornton}, {Van Lanen}, {Vavagiakis}, {Ward}, \&
  {Wollack}}]{Henderson2016}
{Henderson}, S.~W., {Allison}, R., {Austermann}, J., {et~al.} 2016, Journal of
  Low Temperature Physics, 184, 772, \dodoi{10.1007/s10909-016-1575-z}

\bibitem[{Ho {et~al.}(2017)Ho, Austermann, Beall, Choi, Cothard, Crowley,
  Datta, Devlin, Duff, Gallardo, Hasselfield, Henderson, Hilton, Hubmayr,
  Koopman, Li, McMahon, Niemack, Salatino, Simon, Staggs, Ward, Ullom,
  Vavagiakis, \& Wollack}]{Ho2016}
Ho, S.-P.~P., Austermann, J., Beall, J.~A., {et~al.} 2017, in Millimeter,
  Submillimeter, and Far-Infrared Detectors and Instrumentation for Astronomy
  VIII, ed. W.~S. Holland \& J.~Zmuidzinas, Vol. 9914, International Society
  for Optics and Photonics (SPIE), 991418, \dodoi{10.1117/12.2233113}

\bibitem[{{Johnston} {et~al.}(1982){Johnston}, {Seidelmann}, \&
  {Wade}}]{Johnston1982}
{Johnston}, K.~J., {Seidelmann}, P.~K., \& {Wade}, C.~M. 1982, \aj, 87, 1593,
  \dodoi{10.1086/113249}

\bibitem[{{Keihm} {et~al.}(2013){Keihm}, {Kamp}, {Gulkis}, {Hofstadter}, {Lee},
  {Janssen}, \& {Choukroun}}]{Keihm2013}
{Keihm}, S., {Kamp}, L., {Gulkis}, S., {et~al.} 2013, \icarus, 226, 1086,
  \dodoi{10.1016/j.icarus.2013.07.005}

\bibitem[{{Lebofsky} {et~al.}(1986){Lebofsky}, {Sykes}, {Tedesco}, {Veeder},
  {Matson}, {Brown}, {Gradie}, {Feierberg}, \& {Rudy}}]{Lebofsky1986}
{Lebofsky}, L.~A., {Sykes}, M.~V., {Tedesco}, E.~F., {et~al.} 1986, \icarus,
  68, 239, \dodoi{10.1016/0019-1035(86)90021-7}

\bibitem[{Li {et~al.}(2021)Li, Austermann, Beall, Bruno, Choi, Cothard,
  Crowley, Duff, Ho, Golec, Hilton, Hasselfield, Hubmayr, Koopman, Lungu,
  McMahon, Niemack, Page, Salatino, Simon, Staggs, Stevens, Ullom, Vavagiakis,
  Wang, Wollack, \& Xu}]{Li2021}
Li, Y., Austermann, J.~E., Beall, J.~A., {et~al.} 2021, IEEE Transactions on
  Applied Superconductivity, 31, 1, \dodoi{10.1109/TASC.2021.3063334}

\bibitem[{{Li} {et~al.}(2023){Li}, {Biermann}, {Naess}, {Aiola}, {An},
  {Battaglia}, {Bhandarkar}, {Calabrese}, {Choi}, {Crowley}, {Devlin}, {Duell},
  {Duff}, {Dunkley}, {Dunner}, {Gallardo}, {Guan}, {Hervias-Caimapo}, {Hincks},
  {Hubmayr}, {Huffenberger}, {Hughes}, {Kosowsky}, {Louis}, {Mallaby-Kay},
  {McMahon}, {Nati}, {Niemack}, {Orlowski-Scherer}, {Page}, {Sifon},
  {Salatino}, {Staggs}, {Vargas}, {Vavagiakis}, {Wang}, \& {Wollack}}]{Li2023}
{Li}, Y., {Biermann}, E., {Naess}, S., {et~al.} 2023, arXiv e-prints,
  arXiv:2303.04767, \dodoi{10.48550/arXiv.2303.04767}

\bibitem[{{Mainzer} {et~al.}(2011){Mainzer}, {Bauer}, {Grav}, {Masiero},
  {Cutri}, {Dailey}, {Eisenhardt}, {McMillan}, {Wright}, {Walker}, {Jedicke},
  {Spahr}, {Tholen}, {Alles}, {Beck}, {Brandenburg}, {Conrow}, {Evans},
  {Fowler}, {Jarrett}, {Marsh}, {Masci}, {McCallon}, {Wheelock}, {Wittman},
  {Wyatt}, {DeBaun}, {Elliott}, {Elsbury}, {Gautier}, {Gomillion}, {Leisawitz},
  {Maleszewski}, {Micheli}, \& {Wilkins}}]{Mainzer2011}
{Mainzer}, A., {Bauer}, J., {Grav}, T., {et~al.} 2011, \apj, 731, 53,
  \dodoi{10.1088/0004-637X/731/1/53}

\bibitem[{{Marsden} {et~al.}(2014){Marsden}, {Gralla}, {Marriage}, {Switzer},
  {Partridge}, {Massardi}, {Morales}, {Addison}, {Bond}, {Crichton}, {Das},
  {Devlin}, {D{\"u}nner}, {Hajian}, {Hilton}, {Hincks}, {Hughes}, {Irwin},
  {Kosowsky}, {Menanteau}, {Moodley}, {Niemack}, {Page}, {Reese}, {Schmitt},
  {Sehgal}, {Sievers}, {Staggs}, {Swetz}, {Thornton}, \&
  {Wollack}}]{Marsden2014}
{Marsden}, D., {Gralla}, M., {Marriage}, T.~A., {et~al.} 2014, \mnras, 439,
  1556, \dodoi{10.1093/mnras/stu001}

\bibitem[{{Marsset} {et~al.}(2017){Marsset}, {Carry}, {Dumas}, {Hanu{\v{s}}},
  {Viikinkoski}, {Vernazza}, {M{\"u}ller}, {Delbo}, {Jehin}, {Gillon}, {Grice},
  {Yang}, {Fusco}, {Berthier}, {Sonnett}, {Kugel}, {Caron}, \&
  {Behrend}}]{Marsset2017}
{Marsset}, M., {Carry}, B., {Dumas}, C., {et~al.} 2017, \aap, 604, A64,
  \dodoi{10.1051/0004-6361/201731021}

\bibitem[{{Masiero} {et~al.}(2011){Masiero}, {Mainzer}, {Grav}, {Bauer},
  {Cutri}, {Dailey}, {Eisenhardt}, {McMillan}, {Spahr}, {Skrutskie}, {Tholen},
  {Walker}, {Wright}, {DeBaun}, {Elsbury}, {Gautier}, {Gomillion}, \&
  {Wilkins}}]{Masiero2011}
{Masiero}, J.~R., {Mainzer}, A.~K., {Grav}, T., {et~al.} 2011, \apj, 741, 68,
  \dodoi{10.1088/0004-637X/741/2/68}

\bibitem[{{Michel} {et~al.}(2015){Michel}, {DeMeo}, \& {Bottke}}]{Michel2015}
{Michel}, P., {DeMeo}, F.~E., \& {Bottke}, W.~F. 2015, in Asteroids IV
  (University of Arizona Press), 3--10,
  \dodoi{10.2458/azu_uapress_9780816532131-ch001}

\bibitem[{{Moeyens} {et~al.}(2020){Moeyens}, {Myhrvold}, \&
  {Ivezi{\'c}}}]{Moeyens2020}
{Moeyens}, J., {Myhrvold}, N., \& {Ivezi{\'c}}, {\v{Z}}. 2020, \icarus, 341,
  113575, \dodoi{10.1016/j.icarus.2019.113575}

\bibitem[{{Mommert} {et~al.}(2018){Mommert}, {Jedicke}, \&
  {Trilling}}]{Mommert2018}
{Mommert}, M., {Jedicke}, R., \& {Trilling}, D.~E. 2018, \aj, 155, 74,
  \dodoi{10.3847/1538-3881/aaa23b}

\bibitem[{{M{\"u}ller} \& {Barnes}(2007)}]{Muller2007}
{M{\"u}ller}, T.~G., \& {Barnes}, P.~J. 2007, \aap, 467, 737,
  \dodoi{10.1051/0004-6361:20066626}

\bibitem[{{Naess} {et~al.}(2020){Naess}, {Aiola}, {Austermann}, {Battaglia},
  {Beall}, {Becker}, {Bond}, {Calabrese}, {Choi}, {Cothard}, {Crowley},
  {Darwish}, {Datta}, {Denison}, {Devlin}, {Duell}, {Duff}, {Duivenvoorden},
  {Dunkley}, {D{\"u}nner}, {Fox}, {Gallardo}, {Halpern}, {Han}, {Hasselfield},
  {Hill}, {Hilton}, {Hilton}, {Hincks}, {Hlo{\v{z}}ek}, {Ho}, {Hubmayr},
  {Huffenberger}, {Hughes}, {Kosowsky}, {Louis}, {Madhavacheril}, {McMahon},
  {Moodley}, {Nati}, {Nibarger}, {Niemack}, {Page}, {Partridge}, {Salatino},
  {Schaan}, {Schillaci}, {Schmitt}, {Sherwin}, {Sehgal}, {Sif{\'o}n},
  {Spergel}, {Staggs}, {Stevens}, {Storer}, {Ullom}, {Vale}, {Van Engelen},
  {Van Lanen}, {Vavagiakis}, {Wollack}, \& {Xu}}]{Naess2020}
{Naess}, S., {Aiola}, S., {Austermann}, J.~E., {et~al.} 2020, \jcap, 2020, 046,
  \dodoi{10.1088/1475-7516/2020/12/046}

\bibitem[{{Page} {et~al.}(2003){Page}, {Jackson}, {Barnes}, {Bennett},
  {Halpern}, {Hinshaw}, {Jarosik}, {Kogut}, {Limon}, {Meyer}, {Spergel},
  {Tucker}, {Wilkinson}, {Wollack}, \& {Wright}}]{Page2003}
{Page}, L., {Jackson}, C., {Barnes}, C., {et~al.} 2003, \apj, 585, 566,
  \dodoi{10.1086/346078}

\bibitem[{Parshley {et~al.}(2018)Parshley, Niemack, Hills, Dicker, Dünner,
  Erler, Gallardo, Gudmundsson, Herter, Koopman, Limon, Matsuda, Mauskopf,
  Riechers, Stacey, \& Vavagiakis}]{Parshley2018}
Parshley, S.~C., Niemack, M., Hills, R., {et~al.} 2018, in 2018SPIE10700, ed.
  H.~K. Marshall \& J.~Spyromilio, Vol. 10700, International Society for Optics
  and Photonics (SPIE), 1292 -- 1304, \dodoi{10.1117/12.2314073}

\bibitem[{{Planck Collaboration} {et~al.}(2014){Planck Collaboration}, {Ade},
  {Aghanim}, {Armitage-Caplan}, {Arnaud}, {Ashdown}, {Atrio-Barandela},
  {Aumont}, {Baccigalupi}, {Banday}, {Barreiro}, {Bartlett}, {Battaner},
  {Benabed}, {Beno{\^\i}t}, {Benoit-L{\'e}vy}, {Bernard}, {Bersanelli},
  {Bielewicz}, {Bobin}, {Bock}, {Bonaldi}, {Bond}, {Borrill}, {Bouchet},
  {Boulanger}, {Bridges}, {Bucher}, {Burigana}, {Butler}, {Cardoso},
  {Catalano}, {Chamballu}, {Chary}, {Chen}, {Chiang}, {Chiang}, {Christensen},
  {Church}, {Clements}, {Colley}, {Colombi}, {Colombo}, {Couchot}, {Coulais},
  {Crill}, {Curto}, {Cuttaia}, {Danese}, {Davies}, {de Bernardis}, {de Rosa},
  {de Zotti}, {Delabrouille}, {Delouis}, {D{\'e}sert}, {Dickinson}, {Diego},
  {Dole}, {Donzelli}, {Dor{\'e}}, {Douspis}, {Dupac}, {Efstathiou},
  {En{\ss}lin}, {Eriksen}, {Finelli}, {Forni}, {Frailis}, {Fraisse},
  {Franceschi}, {Galeotta}, {Ganga}, {Giard}, {Giraud-H{\'e}raud},
  {Gonz{\'a}lez-Nuevo}, {G{\'o}rski}, {Gratton}, {Gregorio}, {Gruppuso},
  {Hansen}, {Hanson}, {Harrison}, {Helou}, {Henrot-Versill{\'e}},
  {Hern{\'a}ndez-Monteagudo}, {Herranz}, {Hildebrandt}, {Hivon}, {Hobson},
  {Holmes}, {Hornstrup}, {Hovest}, {Huffenberger}, {Jaffe}, {Jaffe}, {Jones},
  {Juvela}, {Keih{\"a}nen}, {Keskitalo}, {Kisner}, {Kneissl}, {Knoche}, {Knox},
  {Kunz}, {Kurki-Suonio}, {Lagache}, {L{\"a}hteenm{\"a}ki}, {Lamarre},
  {Lasenby}, {Laureijs}, {Lawrence}, {Leonardi}, {Lesgourgues}, {Liguori},
  {Lilje}, {Linden-V{\o}rnle}, {L{\'o}pez-Caniego}, {Lubin},
  {Mac{\'\i}as-P{\'e}rez}, {Maffei}, {Maino}, {Mandolesi}, {Maris}, {Marshall},
  {Martin}, {Mart{\'\i}nez-Gonz{\'a}lez}, {Masi}, {Massardi}, {Matarrese},
  {Matthai}, {Mazzotta}, {Meinhold}, {Melchiorri}, {Mendes}, {Mennella},
  {Migliaccio}, {Mitra}, {Miville-Desch{\^e}nes}, {Moneti}, {Montier},
  {Morgante}, {Mortlock}, {Mottet}, {Munshi}, {Murphy}, {Naselsky}, {Nati},
  {Natoli}, {Netterfield}, {N{\o}rgaard-Nielsen}, {Noviello}, {Novikov},
  {Novikov}, {Osborne}, {O'Sullivan}, {Oxborrow}, {Paci}, {Pagano}, {Pajot},
  {Paladini}, {Paoletti}, {Pasian}, {Patanchon}, {Perdereau}, {Perotto},
  {Perrotta}, {Piacentini}, {Piat}, {Pierpaoli}, {Pietrobon}, {Plaszczynski},
  {Pointecouteau}, {Polegre}, {Polenta}, {Ponthieu}, {Popa}, {Poutanen},
  {Pratt}, {Pr{\'e}zeau}, {Prunet}, {Puget}, {Rachen}, {Reach}, {Rebolo},
  {Reinecke}, {Remazeilles}, {Renault}, {Ricciardi}, {Riller}, {Ristorcelli},
  {Rocha}, {Rosset}, {Roudier}, {Rowan-Robinson}, {Rusholme}, {Sandri},
  {Santos}, {Savini}, {Scott}, {Seiffert}, {Shellard}, {Smoot}, {Spencer},
  {Starck}, {Stolyarov}, {Stompor}, {Sudiwala}, {Sureau}, {Sutton},
  {Suur-Uski}, {Sygnet}, {Tauber}, {Tavagnacco}, {Terenzi}, {Toffolatti},
  {Tomasi}, {Tristram}, {Tucci}, {Tuovinen}, {Umana}, {Valenziano},
  {Valiviita}, {Van Tent}, {Vielva}, {Villa}, {Vittorio}, {Wade}, {Wandelt},
  {Yvon}, {Zacchei}, \& {Zonca}}]{Planck2014XIV}
{Planck Collaboration}, {Ade}, P.~A.~R., {Aghanim}, N., {et~al.} 2014, \aap,
  571, A14, \dodoi{10.1051/0004-6361/201321562}

\bibitem[{{Redman} {et~al.}(1992){Redman}, {Feldman}, {Matthews}, {Halliday},
  \& {Creutzberg}}]{Redman1992}
{Redman}, R.~O., {Feldman}, P.~A., {Matthews}, H.~E., {Halliday}, I., \&
  {Creutzberg}, F. 1992, \aj, 104, 405, \dodoi{10.1086/116248}

\bibitem[{{Swetz} {et~al.}(2011){Swetz}, {Ade}, {Amiri}, {Appel},
  {Battistelli}, {Burger}, {Chervenak}, {Devlin}, {Dicker}, {Doriese},
  {D{\"u}nner}, {Essinger-Hileman}, {Fisher}, {Fowler}, {Halpern},
  {Hasselfield}, {Hilton}, {Hincks}, {Irwin}, {Jarosik}, {Kaul}, {Klein},
  {Lau}, {Limon}, {Marriage}, {Marsden}, {Martocci}, {Mauskopf}, {Moseley},
  {Netterfield}, {Niemack}, {Nolta}, {Page}, {Parker}, {Staggs}, {Stryzak},
  {Switzer}, {Thornton}, {Tucker}, {Wollack}, \& {Zhao}}]{Swetz2011}
{Swetz}, D.~S., {Ade}, P.~A.~R., {Amiri}, M., {et~al.} 2011, \apjs, 194, 41,
  \dodoi{10.1088/0067-0049/194/2/41}

\bibitem[{{The CMB-HD Collaboration} {et~al.}(2022){The CMB-HD Collaboration},
  {:}, {Aiola}, {Akrami}, {Basu}, {Boylan-Kolchin}, {Brinckmann}, {Bryan},
  {Casey}, {Chluba}, {Clesse}, {Cyr-Racine}, {Di Mascolo}, {Dicker},
  {Essinger-Hileman}, {Farren}, {Fedderke}, {Ferraro}, {Fuller}, {Galitzki},
  {Gluscevic}, {Grin}, {Han}, {Hasselfield}, {Hlozek}, {Holder}, {Hotinli},
  {Jain}, {Johnson}, {Johnson}, {Klaassen}, {MacInnis}, {Madhavacheril},
  {Mandal}, {Mauskopf}, {Meerburg}, {Meyers}, {Miranda}, {Mroczkowski},
  {Mukherjee}, {Munchmeyer}, {Munoz}, {Naess}, {Nagai}, {Namikawa}, {Newburgh},
  {Nguyen}, {Niemack}, {Oppenheimer}, {Pierpaoli}, {Raghunathan}, {Schaan},
  {Sehgal}, {Sherwin}, {Simon}, {Slosar}, {Smith}, {Spergel}, {Switzer},
  {Trivedi}, {Tsai}, {van Engelen}, {Wandelt}, {Wollack}, \& {Wu}}]{CMBHD2022}
{The CMB-HD Collaboration}, {:}, {Aiola}, S., {et~al.} 2022, arXiv e-prints,
  arXiv:2203.05728, \dodoi{10.48550/arXiv.2203.05728}

\bibitem[{{Tholen}(1984)}]{Tholen1984}
{Tholen}, D.~J. 1984, PhD thesis, University of Arizona

\bibitem[{{Thornton} {et~al.}(2016){Thornton}, {Ade}, {Aiola}, {Angil{\`e}},
  {Amiri}, {Beall}, {Becker}, {Cho}, {Choi}, {Corlies}, {Coughlin}, {Datta},
  {Devlin}, {Dicker}, {D{\"u}nner}, {Fowler}, {Fox}, {Gallardo}, {Gao},
  {Grace}, {Halpern}, {Hasselfield}, {Henderson}, {Hilton}, {Hincks}, {Ho},
  {Hubmayr}, {Irwin}, {Klein}, {Koopman}, {Li}, {Louis}, {Lungu}, {Maurin},
  {McMahon}, {Munson}, {Naess}, {Nati}, {Newburgh}, {Nibarger}, {Niemack},
  {Niraula}, {Nolta}, {Page}, {Pappas}, {Schillaci}, {Schmitt}, {Sehgal},
  {Sievers}, {Simon}, {Staggs}, {Tucker}, {Uehara}, {van Lanen}, {Ward}, \&
  {Wollack}}]{Thornton2016}
{Thornton}, R.~J., {Ade}, P.~A.~R., {Aiola}, S., {et~al.} 2016, \apjs, 227, 21,
  \dodoi{10.3847/1538-4365/227/2/21}

\bibitem[{{Viikinkoski} {et~al.}(2015){Viikinkoski}, {Kaasalainen},
  {{\v{D}}urech}, {Carry}, {Marsset}, {Fusco}, {Dumas}, {Merline}, {Yang},
  {Berthier}, {Kervella}, \& {Vernazza}}]{Viikinkoski2015}
{Viikinkoski}, M., {Kaasalainen}, M., {{\v{D}}urech}, J., {et~al.} 2015, \aap,
  581, L3, \dodoi{10.1051/0004-6361/201526626}

\bibitem[{{Webster} {et~al.}(1988){Webster}, {Johnston}, {Hobbs}, {Lamphear},
  {Wade}, {Lowman}, {Kaplan}, \& {Seidelmann}}]{Webster1988}
{Webster}, W.~J., {Johnston}, K.~J., {Hobbs}, R.~W., {et~al.} 1988, \aj, 95,
  1263, \dodoi{10.1086/114722}

\bibitem[{{Wright} {et~al.}(2010){Wright}, {Eisenhardt}, {Mainzer}, {Ressler},
  {Cutri}, {Jarrett}, {Kirkpatrick}, {Padgett}, {McMillan}, {Skrutskie},
  {Stanford}, {Cohen}, {Walker}, {Mather}, {Leisawitz}, {Gautier}, {McLean},
  {Benford}, {Lonsdale}, {Blain}, {Mendez}, {Irace}, {Duval}, {Liu}, {Royer},
  {Heinrichsen}, {Howard}, {Shannon}, {Kendall}, {Walsh}, {Larsen}, {Cardon},
  {Schick}, {Schwalm}, {Abid}, {Fabinsky}, {Naes}, \& {Tsai}}]{Wright2010}
{Wright}, E.~L., {Eisenhardt}, P. R.~M., {Mainzer}, A.~K., {et~al.} 2010, \aj,
  140, 1868, \dodoi{10.1088/0004-6256/140/6/1868}

\bibitem[{{Zhu} {et~al.}(2021){Zhu}, {Bhandarkar}, {Coppi}, {Kofman},
  {Orlowski-Scherer}, {Xu}, {Adachi}, {Ade}, {Aiola}, {Austermann}, {Bazarko},
  {Beall}, {Bhimani}, {Bond}, {Chesmore}, {Choi}, {Connors}, {Cothard},
  {Devlin}, {Dicker}, {Dober}, {Duell}, {Duff}, {D{\"u}nner}, {Fabbian},
  {Galitzki}, {Gallardo}, {Golec}, {Haridas}, {Harrington}, {Healy}, {Ho},
  {Huber}, {Hubmayr}, {Iuliano}, {Johnson}, {Keating}, {Kiuchi}, {Koopman},
  {Lashner}, {Lee}, {Li}, {Limon}, {Link}, {Lucas}, {McCarrick}, {Moore},
  {Nati}, {Newburgh}, {Niemack}, {Pierpaoli}, {Randall}, {Sarmiento},
  {Saunders}, {Seibert}, {Sierra}, {Sonka}, {Spisak}, {Sutariya}, {Tajima},
  {Teply}, {Thornton}, {Tsan}, {Tucker}, {Ullom}, {Vavagiakis}, {Vissers},
  {Walker}, {Westbrook}, {Wollack}, \& {Zannoni}}]{Zhu2021}
{Zhu}, N., {Bhandarkar}, T., {Coppi}, G., {et~al.} 2021, \apjs, 256, 23,
  \dodoi{10.3847/1538-4365/ac0db7}

\end{thebibliography}
\bibliographystyle{aasjournal}

\end{document}